\documentclass[aps,physrev,twocolumn,groupedaddress]{revtex4-2}

\usepackage{amssymb}
\usepackage[disable]{todonotes}
\usepackage[colorlinks=true,allcolors=blue,breaklinks=true]{hyperref}
\usepackage{lineno}
\begin{document}


\title{
Tracing complex zeros of~the~quantum survival amplitude: How the~energy distribution controls dynamical phase transitions
}


\author{Jakub Novotn\'{y}}
\email[]{jakub.novotny@matfyz.cuni.cz}
\affiliation{Institute of~Particle and Nuclear Physics, Faculty of~Mathematics and Physics,
Charles University, V Hole\v{s}ovi\v{c}k\'{a}ch 2, 18000 Prague, Czech Republic}


\author{Jan St\v{r}ele\v{c}ek}


\email[]{jan.strelecek@matfyz.cuni.cz}
\affiliation{Institute of~Particle and Nuclear Physics, Faculty of~Mathematics and Physics,
Charles University, V Hole\v{s}ovi\v{c}k\'{a}ch 2, 18000 Prague, Czech Republic}

\author{Pavel Str\'{a}nsk\'{y}}


\affiliation{Institute of~Particle and Nuclear Physics, Faculty of~Mathematics and Physics,
Charles University, V Hole\v{s}ovi\v{c}k\'{a}ch 2, 18000 Prague, Czech Republic}
\author{Pavel Cejnar}


\affiliation{Institute of~Particle and Nuclear Physics, Faculty of~Mathematics and Physics,
Charles University, V Hole\v{s}ovi\v{c}k\'{a}ch 2, 18000 Prague, Czech Republic}


\date{\today}

\begin{abstract}
Motivated by the~advance of~dynamical quantum phase transitions (DQPTs), we analyze the~zeros of~the~complex-time survival (Loschmidt) amplitude in~finite quantum systems and develop a~general framework for their approximation based on the~stability of~zeros of~holomorphic functions.
We show that the~large-scale properties of~the~distribution of~zeros are governed by the~envelope of~the~energy distribution of~the~initial state and can be constructed from chains of~periodic zeros associated with its dominant contributions.
In this picture, zeros reach the~real-time axis when two or more eigenstates become equally populated at the~maximum of~the~envelope, providing a~finite-size precursor of~DQPTs.
We apply the~method to quenched ground states in~the~Ising model with tunable interaction range and demonstrate close agreement between the~approximate and exact distributions of~zeros.
We prove that the~approximate construction becomes exact for BCS ground-state quenches in~two-band models. To describe short-time dynamics, we introduce a~minimal Gaussian model with a~nearly equidistant spectrum. Slow dephasing continuously deforms the~initial zero pattern into the~asymptotic two-level structure, explaining anomalous DQPTs as a~delayed approach of~zeros to the~real-time axis.
Our results identify the~energy envelope as the~key ingredient shaping dynamical critical behavior and provide a~universal interpretation of~the~whole zero distribution of~the~complex-time survival amplitude.
\end{abstract}


\maketitle

\section{Introduction \label{Introduction}}

The survival amplitude, defined as the~overlap between the~initial quantum state and its time-evolved counterpart, 
has attracted attention in~a~new context due to its formal resemblance to the~partition function. 
In equilibrium thermodynamics, the~partition function plays a~central role, and its zeros are associated with phase transitions~\cite{YangLee,Fisher1965,bena2005statistical}.
In the~thermodynamic limit, these zeros generate singularities in~the~logarithm of~the~partition function, which corresponds to the thermodynamic potential.
The analogy raises the~question whether the~zeros of~the~survival amplitude (also called the~Loschmidt amplitude) indicate a~transition between time domains with distinct dynamical behavior, analogous to equilibrium phase transitions.

This motivated the~term Dynamical Quantum Phase Transition (DQPT)~\cite{heyl2013dynamical,heyl2018dynamical} for zeros of~the~survival amplitude on the~real-time axis in~the~thermodynamic limit. 
The standard setting for studying DQPTs
\footnote{Also sometimes called the~DQPT-II to distinguish it from a~different concept of~DQPT(-I)~\cite{DQPT-I_Halimeh,marino2022dynamical}} became the~quench of~the~ground state for some initial parameters.
In the~pioneering work~\cite{heyl2013dynamical}, the~authors showed that for the~one-dimensional transverse-field Ising model, zeros appeared on the~real-time axis only for quenches across the~underlying Quantum Phase Transition (QPT), signaling a~connection between DQPTs and equilibrium criticality.
In those cases, the~zeros formed a~periodic pattern along the~real-time axis.
Analytic continuation to complex times revealed lines of~zeros in~the~thermodynamic limit that separated regions of~the~complex-time plane, in~direct analogy to Fisher zeros in~equilibrium phase transitions.

The periodic patterns of~lines of~zeros are characteristic of~two-band models of~noninteracting particles in~1D
\cite{PhysRevB.89.161105,PhysRevB.91.155127,PhysRevB.99.121107,PhysRevB.94.064423,PhysRevB.99.054302,PhysRevLett.122.050403,PhysRevB.103.064306,PhysRevB.105.094514}
and these lines generalize to regions of~zeros in~2D systems
\cite{schmitt2015dynamical,PhysRevB.95.1843072Dtopology,PhysRevB.108.094306,PhysRevB.108.094306,Maslowski_2024,PhysRevB.109.134301}.
In this context, the~connection between DQPTs and QPTs often arises in~topological systems 
\cite{PhysRevB.91.155127,huang2016dynamical,PhysRevB.96.014302,PhysRevResearch.1.033039,Flaschner2018,PhysRevResearch.2.033259,PhysRevResearch.3.043064,PhysRevB.106.224302}. 
The~appearance of~zeros on the~real-time axis can be explained by crossing a~topological transition during the~quench, and in~some cases, by introducing a~dynamical order parameter~\cite{PhysRevB.93.085416,PhysRevB.109.L140303}.

However, subsequent work showed that these features are not generic.
The direct correspondence between DQPTs and QPTs is violated even for some noninteracting models~\cite{PhysRevB.89.161105,RYLANDS2021168554} and breaks down with the~introduction of~an interaction~\cite{PhysRevB.89.125120,PhysRevB.89.054301,anomalous2017,anomalousII2017,PhysRevLett.120.130601,RYLANDS2021168554,PhysRevResearch.7.023194}.

DQPTs often occur when a~quench crosses an equilibrium quantum critical point even in~interacting systems~\cite{PRBKarrasch2013,PhysRevB.90.125106,PhysRevB.103.064306}.
However, zeros of~the~survival amplitude may also reach the~real-time axis after quenches that do not cross any quantum critical point, indicating that their presence is not uniquely tied to the~equilibrium phase structure.
Moreover, introducing interactions, disorder, or going beyond two-band models, generally disturbs the~periodic structure of~zeros in~time 
\cite{PRBKarrasch2013,PhysRevB.90.125106, PhysRevB.92.104306, PhysRevLett.122.250601, PhysRevB.104.085104, PhysRevB.101.014305,PhysRevB.100.224307,PhysRevLett.122.250601,Mishra_2020, PhysRevB.101.014301,RYLANDS2021168554,PhysRevB.107.094307}
and may smooth the~nonanalyticities in~the~logarithm of~the~survival probability, i.e., the~rate function. 
This includes scenarios in~which sharp nonanalyticities develop only at long times
and are preceded by smooth temporal oscillations, a~phenomenon often referred to as Anomalous Dynamical Quantum Phase Transitions (ADQPT)~\cite{anomalous2017,anomalousII2017,PhysRevLett.120.130601,PhysRevE.96.062118,PhysRevResearch.2.033111,PhysRevB.100.014434,PhysRevResearch.4.013250,Niu_2024}.
While such behavior has been observed, for example, in~the~Ising model with long-range interaction, it is not a~generic consequence of~interaction range alone~\cite{anomalousisnotlongrange}.

Although DQPTs and zeros of the survival amplitude are typically interpreted within specific models or theoretical frameworks---such as topological settings, transfer-matrix approaches~\cite{PhysRevB.89.125120,PhysRevLett.126.040602,PhysRevB.105.165149}, or analytical treatments of~DQPTs~\cite{PhysRevX.11.041018,PhysRevResearch.4.033032,PhysRevB.100.035124}---a unifying physical interpretation is still lacking.
Nevertheless, the~DQPTs have attracted experimental interest~\cite{PhysRevLett.119.080501,Flaschner2018,PhysRevB.100.024310,PhysRevApplied.11.044080,PhysRevLett.122.020501,xu2020measuring,Dborin2022},
and have been extended to various settings, including finite-temperature~\cite{PhysRevB.93.104302,PhysRevB.97.045147}, driven~\cite{PhysRevA.97.053621,PhysRevB.100.085308}, and non-Hermitian systems~\cite{PhysRevA.98.022129}.
 
In this work, we focus on zeros of~the~complex-time survival amplitude in~a~general finite-dimensional closed quantum system.
These zeros can be viewed as finite-size precursors of~DQPTs. 
We introduce a~general method to approximate the~exact distribution of~zeros.
Our approach exploits the~stability of~zeros of~holomorphic functions under perturbations, allowing approximate distributions of~zeros to be constructed from dominant contributions to the~survival amplitude.
Rather than resolving the~exact positions of~individual zeros, the~method is designed to capture the~large-scale structure of~their distribution.
This allows us to relate the~organization of~zeros in~the~complex-time plane to the~envelope of~the~initial-state energy distribution, also known as the~local density of~states or strength function.
Changes of~the~initial state or the~system parameters can then be interpreted through the~corresponding deformation of~this envelope, providing a~general way to discuss zeros of~the~survival amplitude beyond model-specific mechanisms.

This paper is organized as follows.
In Sec.~\ref{sec:II}, we define the~energy envelope of~the~initial-state energy distribution and develop a~general approximation scheme for the~zeros of~the~complex-time survival amplitude, based on the~local dominance of~individual contributions and the~stability of~zeros of~holomorphic functions.
In Sec.~\ref{sec:III}, we apply this framework to quenched ground states of~the~1D transverse-field Ising model with tunable interaction range, where we compare approximate and exact zero distributions and clarify when zeros approach the~real-time axis based on the~shape of~the~envelope.
We prove that the~construction is exact for BCS ground-state quenches in~two-band models and analyze how the~quench length affects the~distribution of~zeros near the~real-time axis in~the~collective long-range limit.
In Sec.~\ref{sec:IV}, we focus on systems with nearly equidistant spectra and introduce a~minimal Gaussian model to analyze the~short-time distribution of~zeros, the~role of~slow dephasing, and the~resulting delayed approach of~zeros to the~real-time axis associated with ADQPTs.
In Sec.~\ref{sec:V}, we summarize the~results and discuss implications for finite-size precursors of~DQPTs.

\section{Approximate distribution of~zeros from the~energy envelope}\label{sec:II}
We consider a~system described by a~finite-dimensional Hamiltonian $\hat{H}$, with a~discrete spectrum $\{E_j\}_{j=0}^d$ and eigenvectors $\{|E_j\rangle\}_{j=0}^d$, in~an initial state $|\psi\rangle$ defined by the~energy distribution
\begin{equation}
k_{\psi}(E) = \sum_{j=0}^d k_j~\delta(E-E_j),
\end{equation}
with probabilities $k_j = |\langle\psi|E_j\rangle|^2$.
The complex-time survival amplitude is a~Laplace transform of~the~energy distribution $k_{\psi}(E)$, and for the~finite system it corresponds to a~weighted sum of~complex exponentials
\begin{equation}
    \label{def:overlapComplex}
    \mathcal{L}(z)
= \int k_{\psi}(E)e^{-Ez}\,dE
= \sum_{j=0}^d \underbrace{k_j e^{-E_j z}}_{\ell_j(z)},
\end{equation}
where the~physical time $t$ is the~imaginary part of~the~complex variable
\begin{equation}\label{eq:complexVariable}
    z = \beta + it.
\end{equation}
The real part $\beta$ plays a~role similar to the~inverse temperature.
By changing $\beta$, we effectively shift the~energy distribution to higher or lower energies. 
In what follows, we refer to the~line $\beta=0$ as the~real-time axis, since it corresponds to physical real-time evolution.

Cast in~polar form, the~radial part $r_j$ of~each term $\ell_j$ depends solely on $\beta$ while the~phase $\phi_j$ is a~function of~time $t$,
\begin{equation}
    \ell_j = k_j e^{-E_j \beta} e^{-iE_jt } = r_j(\beta) e^{i \phi_j(t)}.
\end{equation}

A zero, $\mathcal{L}(z) =0$, forms at a~point $z$ in~the~complex plane when these terms $\ell_j$ align to form a~closed polygon~\cite{huang2016dynamical}.
As a~zero of~a~holomorphic function~\cite{Stein2003PrincetonLI}, it is stable under perturbations:
neglecting or adding terms, and thus breaking the~polygon condition, does not eliminate the~zero but shifts it to a~new position in~the~complex plane.

We exploit this stability to construct an approximate distribution of~zeros.
Locally, we reduce the~survival amplitude to its dominant terms and take the~zeros of~the~reduced amplitude as approximate zeros.
Our aim is to construct an approximate distribution that correctly captures the~density of~zeros on coarse scales in~the~complex plane, i.e., the~number of~zeros within a~given region, rather than the exact positions of the individual zeros. 
This allows us to identify which features of the initial state organize the zeros and control how their distribution changes when the state is varied.

In this section, we focus on the~generic case of~non-equidistant spectra.
The basic building blocks of~this distribution are zeros of~effective two-level systems, whose positions can be calculated analytically. 
This distribution can be deformed into the~exact one without creating or annihilating zeros and therefore reproduces the~correct density of~zeros on large scales in~the~complex plane.

\subsection{Schematic example: two- and three-level systems}
\begin{figure}
    \includegraphics[width=\linewidth]{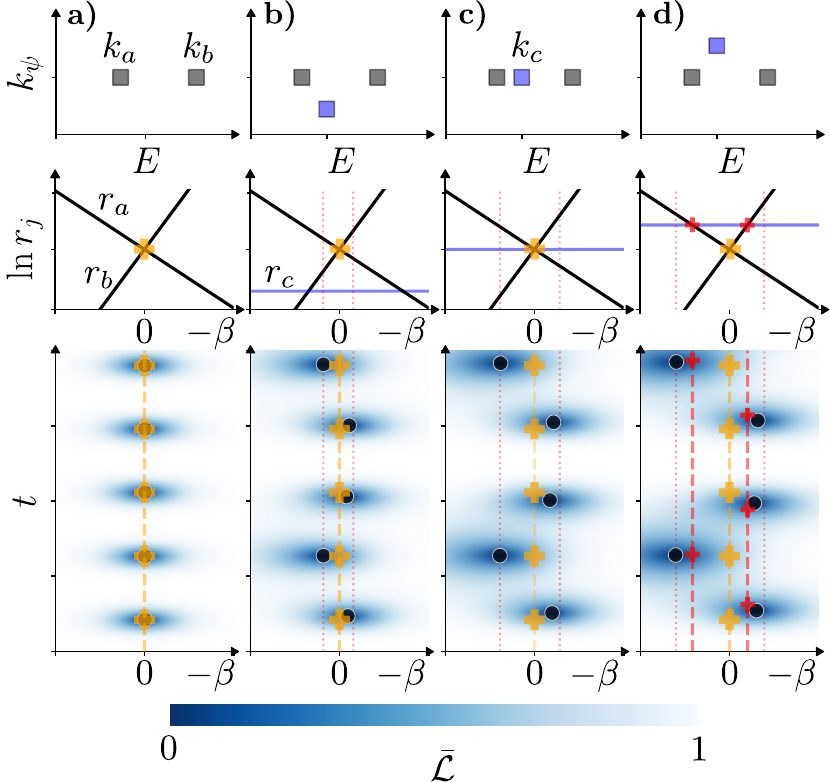}
    \centering
    \caption{\textbf{Top panels:} Energy distribution of~the~initial state (squares) with gradually increasing population $k_c$ from panel (a) to (d) of~the~central level (blue square) set to $E_c = 0$.
    \textbf{Middle panels:} Logarithmic plots of~the~radial parts $r_j(\beta)$.
    Intersections mark cancellations between dominant terms: orange crosses correspond to cancellations of~the~original edge terms $\ell_a$ and $\ell_b$, while red crosses in~panel (d) indicate cancellations of~the~newly dominant terms.
    The~blue line shows the~increasing population of~the~central level as a~function of~$\beta$.
    \textbf{Bottom panels:} Exact zeros in~the~complex plane (black dots), together with the~two-level zero distribution generated by the~edge terms (orange crosses).
    In~panel (d), red crosses mark the~approximate zero distribution formed by two newly emerging zero chains.
    In~both rows, the~thin red dotted lines indicate the~strip within which the~zeros remain confined after the~perturbation according to Rouché's theorem. 
    The~background shows the~normalized survival amplitude
    $\bar{\mathcal{L}}=\mathcal{L}(\beta+it)/\mathcal{L}(\beta)$.
    }
    \label{fig:Deformation}
\end{figure}

We illustrate the~general principles of~the~approximate zero distribution with a~simple example of~a~few-level system.
Consider a~two-level system with an initial state that populates both energy levels.
The survival amplitude then consists of~terms $\ell_a$ and $\ell_b$.
In the~middle panels of~Fig.~\ref{fig:Deformation}, we plot the~radial parts of~these terms as a~function of~$\beta$ on a~logarithmic scale.
Zeros of~the~survival amplitude occur when the~two terms cancel, which happens at the~intersection of~these lines.

\begin{figure*}
    \includegraphics[width=0.75\linewidth]{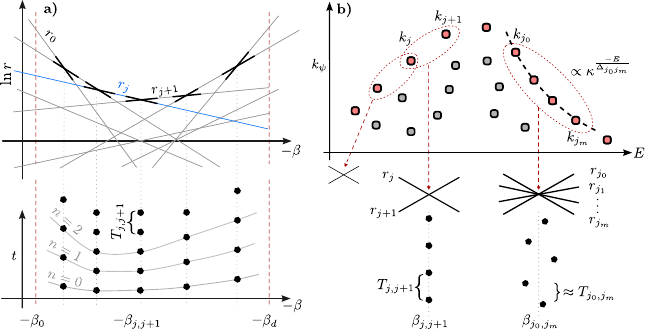}
    \caption{
    \textbf{(a)} Construction of~the~approximate zero distribution from the~crossings of~dominant envelope terms in~the~logarithmic plot of~the~radial terms $r_j$ (upper plot), together with the~corresponding chains of~zeros in~the~complex plane (bottom plot).
    \textbf{(b)} Visualization of~the~energy distribution $k_{\psi}$ of~the~initial state (gray squares) and its envelope (red squares). Consecutive terms generate two-level or multilevel zeros with corresponding two-level or multilevel intersections, depending on the~shape of~the~envelope.
    }
    \label{fig:Theory}
\end{figure*}

The resulting zeros form a~periodic chain located at $\beta_{ab}$ with time period $T_{ab}$ given by 
\begin{equation}
\label{def:betaANDperiod}
 \beta_{ab} =  \frac{1}{\Delta_{ab}} \ln  \left(\frac{k_a}{k_b} \right), \quad T_{ab} =\left|\frac{2\pi}{\Delta_{ab}}\right|, 
\end{equation}
 where $\Delta_{ab} = E_a - E_b$ is the~energy difference between the~two levels.
 The~chain is shown in~the~bottom panel of~Fig.~\ref{fig:Deformation}(a), and we further refer to this pattern of~zeros as \textit{two-level zeros} for brevity.
The individual zeros in~the~chain, enumerated by $n\in\mathbb{Z}$, are located at
\begin{eqnarray}
\label{eq:twilevelZero}
    z_{ab}(n) =& \beta_{ab} + i T_{ab}  \left(n + \frac{1}{2}\right). 
\end{eqnarray}
 
As we populate an additional level by adding a~third term $\ell_c$ in~the~interior of~the~spectrum, $E_a<E_c<E_b$, and gradually increase its weight, the~distribution of~zeros changes. 
In general, the~zero distribution of~a~three-level or multilevel system cannot be described analytically.
Acting as a~holomorphic perturbation, the~additional term gradually deforms the~two-level chain.
According to Rouché's theorem \cite{Stein2003PrincetonLI} (see comment
\footnote{Loosely speaking, when a~holomorphic function $f(z)$ is perturbed by another holomorphic perturbation $g(z)$, the~function $f(z) + g(z)$ has the~same number of~zeros inside any contour on which $|f|>|g|$; see~\cite{Stein2003PrincetonLI}.}),
the zeros cannot be pushed out of~the~strip bounded by the~region in~which one of~the~terms of~the~original two-level system is larger than the~sum of~all the~other terms.
In Figs.~\ref{fig:Deformation}(b)--(d), this corresponds to the~region between the~dotted lines, whose positions are determined by the~conditions $r_a(\beta) = r_b(\beta) + r_c(\beta)$ and $r_b(\beta) = r_a(\beta) + r_c(\beta)$, respectively.
On this scale, the~distribution of~zeros remains well approximated by the~original two-level chain generated by the~edge levels of~the~spectrum.

Once the~weight of~the~additional term $\ell_c$ starts to exceed the~original terms in~the~survival amplitude at a~given $\beta$, the~zero structure becomes locally determined by these newly dominant terms and their cancellations. 
In Figs~\ref{fig:Deformation}(c) and (d), the~original intersection in~the~logarithmic graph splits as $\ell_c$ is increased, producing two new intersections corresponding to the~cancellation of~the~two largest terms.
The resulting local distribution of~zeros is consequently better approximated by the~two new chains of~two-level zeros at $z_{ac}$ and $z_{cb}$ given by Eq.~(\ref{eq:twilevelZero}).  

By populating additional levels between the~edge levels, the~distribution of~zeros continues to deform locally whenever new terms become dominant.
On large scales in~the~complex plane, it remains continuously connected to the~original distribution generated by the~edge terms. 
Therefore, two-level zero chains naturally serve as the~building blocks of~the~approximate zero distribution.

At this point, one may ask why such drastic approximation works.
Although the~two largest terms generally provide a~poor approximation to the~full survival amplitude, they determine the~local zero structure with surprising accuracy.
This follows from the~perturbative stability of~zeros, which can be understood as sources for the~corresponding holomorphic function, analogously to charges in~electrostatics.
We provide the~technical discussion in~Appendix \ref{app:deformation}.

\subsection{General construction for an arbitrary initial state}

We now extend the~construction of~the~approximate zero distribution to a~general initial state of~a~$(d+1)$-level system.
Assuming that the~terms in~the~survival amplitude quickly dephase due to a~far-from-equidistant spectrum, we approximate the~survival amplitude locally by its dominant terms and identify their cancellations with approximate zeros.

For a~given $\beta$, only a~subset of~terms dominates the~survival amplitude.
We define the~\textit{energy envelope} of~the~state as the~set of~terms $\ell_j$ that dominate at some $\beta$, i.e., $\ell_j$  belongs to the~envelope if 
\begin{equation}
\label{def:conditionEnvelope}
    \exists \beta: \quad  r_j(\beta) \geq  r_i(\beta) \quad \forall i.
\end{equation}

In the~logarithmic plot of~the~radial parts, sketched in~Fig.~\ref{fig:Theory}(a), these envelope terms correspond to the~uppermost lines.
The approximate zeros form at the~intersections of~consecutive envelope terms.
This dominance condition 
 can be reformulated explicitly:
 two terms $\ell_a$ and $\ell_b$ with $E_a<E_b$ belong to the~energy envelope of~the~initial state if, at their intersection $\beta_{ab}$, no other term $\ell_c$ is larger, i.e., the~following inequality \footnote{
Note that this inequality is better suited for an algorithmic identification of~the~envelope than the~condition in~Eq.~(\ref{def:conditionEnvelope}).}
holds
\begin{equation}
    \label{eq:inequality}
    k_c^{-\Delta_{ba}} \geq k_a^{\Delta_{cb}}k_b^{\Delta_{ac}}\quad \forall k_c, E_c.
\end{equation}
When the~inequality is satisfied in~the~strict sense,
we approximate the~distribution of~zeros near $\beta_{ab}$ by two-level zeros given by Eq.~(\ref{eq:twilevelZero}).

When more than two dominant terms intersect, the~local structure of~zeros corresponds to an effective multilevel system.
Although the~resulting multilevel zeros cannot, in~general, be obtained analytically, they can be viewed as deformations of~the~two-level zeros generated by the~edge terms.
Specifically, if the~$m+1$ largest terms intersect at $\beta_{j_0j_m}$ (see the~sketch in~Fig.~\ref{fig:Theory}(b)),
the multilevel zeros form around $\beta_{j_0j_m}$ with an approximate period $T_{j_0j_m}$ given by Eq.~(\ref{def:betaANDperiod}).

Thus, the complete approximate zero distribution is constructed from two-level chains of~zeros generated by pairs of~consecutive levels in~the~energy envelope of~the~initial state.
As a result, it is determined solely by the~shape of~the~initial-state energy distribution.
The exact distribution is obtained by deforming these chains through the inclusion of subdominant terms.
Further discussion of~the~effect of~non-leading terms on the~distribution of~zeros is provided in~Appendix~\ref{app:deformation}, together with the~analysis of~multilevel zeros and how they arise from parts of~the~envelope that depend exponentially on energy.

\begin{figure*}
    \includegraphics[width=0.925\linewidth]{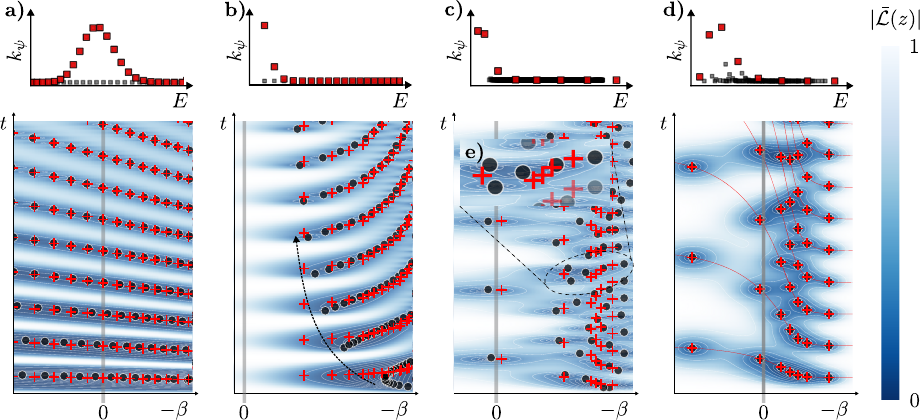}
    \caption{
    \textbf{Top panels:} Energy distributions $k_{\psi}$ of~the~initial states (small dark squares) and their envelopes (red squares).
    \textbf{Bottom panels:} Normalized survival amplitudes $|\bar{\mathcal{L}}|$ plotted in~the~complex plane with exact (black dots) and approximate (red crosses) distributions of~zeros calculated from the~envelopes.
    \textbf{(a)}~Long quench of~the~symmetric ground state from $h_i = 0.2$ to $h_f = 1.5$ in~the~fully connected Ising model ($\alpha = 0$) with $N = 100$.
    \textbf{(b)}~Short quench of~the~symmetric ground state from $h_i = 0.1$ to $h_f = 0.2$ in~the~fully connected Ising model ($\alpha = 0$) with $N = 100$.
    The~dashed arrow indicates the~gradual change of~the~exact zero distribution.
    \textbf{(c)}~Quench of~the~symmetric ground state from $h_i = 0.5$ to $h_f = 1.0$ in~the~long-range Ising model with $\alpha = 1.5$ and $N = 14$.
    Inset~\textbf{(e)} is shown to compare the~exact and approximate distributions on a~finer scale. 
    \textbf{(d)}~Quench of~the~symmetric ground state across the~QPT from $h_i = 0.1$ to $h_f = 0.5$ in~the~nearest-neighbor Ising model ($\alpha = \infty$) and $N = 14$.
Thin red lines mark positions of~coinciding exact and approximate zeros in~the~thermodynamic limit.
    }
    \label{fig:Models}
\end{figure*}

From Eq.~(\ref{eq:inequality}), the~envelope as a~function of~energy can change monotonicity at most once.
The defining feature of~the~envelope---the requirement of~the~dominant terms---smears all oscillations of~the~level distribution of~the~initial state. 
The increasing part of~the~envelope generates zeros in~the~positive-$\beta$ half-plane, while the~decreasing part generates zeros in~the~negative-$\beta$ half-plane,
\begin{equation}
\label{eq:monotonicity}
k_{i+1} \gtreqqless k_{i} \iff \beta_{i+1,i} \gtreqqless 0.
\end{equation}
The distribution of~zeros at the~real-time axis, which corresponds to precursors of~DQPTs, is therefore determined by the~maximum of~the~envelope, where the~monotonicity changes. 
Approximate zeros form precisely at the~real-time axis when the maximum is formed by two or more equally populated states, as shown in~Fig.~\ref{fig:Models}. 

Deformations of~the~energy envelope induced by changes in~system parameters or the~initial state lead directly to deformations of~the~distribution of~zeros.
This provides a~simple way to understand how modifications of~the~initial state—for example, those caused by changing the~quench length—reshape the~distribution of~zeros in~the~complex plane.

\section{Ground state quenches}
\label{sec:III}

We illustrate the~approximate zero distribution and the~role of~the~envelope by considering the~survival amplitude of~a~quenched ground state.
We compare the~approximate and exact distributions in~a~class of~models with well-studied DQPTs~\cite{heyl2013dynamical,anomalous2017,anomalousII2017,PhysRevLett.120.130601,PhysRevB.106.024311,PhysRevB.107.094307}.
As a~representative model, we consider the~one-dimensional Ising model with a~tunable interaction range in~a~transverse magnetic field, with $N$ sites and periodic boundary conditions
\begin{equation}
    \label{LongRangeIsing}
    \hat{H}(h,\alpha) = -\frac{1}{\mathcal{N}} \sum_{i,j}\frac{1}{|i-j|^{\alpha}}\hat{S}_i^z \hat{S}_j^z + h \sum_j\hat{S}_j^x, 
\end{equation}
where $\hat{S}_i$ are the~spin-$\frac{1}{2}$ matrices, $h$ is the~strength of~the~transverse magnetic field, and $\alpha$ sets the~range of~the~interaction.
The factor $\mathcal{N} = \frac{2}{N-1}\sum_n \frac{N-n}{n^{\alpha}}$ is the~Kac normalization~\cite{kac1963van}.
The limiting cases $\alpha = \infty$ and $\alpha = 0$ are the~standard Ising model with nearest-neighbor interaction and the~fully connected case, respectively.
In the~fully connected limit, we restrict the~dynamics to the~largest spin sector, where the~Hamiltonian reduces to the~Lipkin model~\cite{botet1982size}.

Both limiting models are integrable and undergo a~quantum phase transition, the~nearest-neighbor model at $h_c = \frac{1}{4}$, and the~fully connected model at $h_c =1$ in~our parametrization.
In the~nearest-neighbor case, zeros of~the~survival amplitude appear at the~real-time axis only for quenches across the~critical point~\cite{heyl2013dynamical}.
In contrast, for sufficiently long-range interactions ($\alpha \lessapprox 2$), zeros form near the~real-time axis, even for arbitrarily short quenches. 
However, they emerge only at anomalously long times \cite{anomalous2017,anomalousII2017}, rather than at the~first minimum, which is the~characteristic signature of~an ADQPT.

\subsection{Exact versus approximate distributions of~zeros}
We now illustrate how the~envelope of~the~initial-state energy distribution organizes the~zeros of~the~survival amplitude after a~quench.
Fig.~\ref{fig:Models} shows the~distribution of~zeros in~the~complex plane for quenched ground states $|\psi\rangle$ in~Ising models with different interaction ranges $\alpha$.
For each case, we compare the~exact zeros, obtained numerically as described in~Appendix~\ref{app:numerics}, with the~approximate distribution constructed from the~envelope of~$k_{\psi}$.
The comparison shows how changes induced by the~quench are reflected in~the~deformation of~the~envelope and, consequently, in~the~large-scale structure of~the~zero distribution.

Figs~\ref{fig:Models}(a) and \ref{fig:Models}(b)
show results for two quenches of~different lengths in~the~fully connected model ($\alpha = 0$). 
In the~long-quench case, the~envelope of~the~initial state exhibits a~bell-shaped distribution with increasing and decreasing intervals, resulting in~zeros in~both positive- and negative-$\beta$ half-planes.
This is accurately captured by the~approximate distribution, with the~pair of~largest terms in~the~envelope generating the~zeros closest to the~real-time axis.
In the~short-quench case, the~initial state remains close to the~ground state of~the~final Hamiltonian.
The envelope is monotonically decreasing and generates zeros only in~the~negative-$\beta$ half-plane, with no zeros appearing near the~real-time axis.
Here, the~approximate zero distribution deviates from the~exact one at short times.
As indicated by the~dashed arrow, the~distribution of~zeros at short times appears to evolve toward the~approximate distribution.
We directly relate this behavior to the~nearly equidistant structure of~the~envelope and to the~ADQPTs in~Sec.~\ref{sec:IV}.

Fig.~\ref{fig:Models}(c) shows the~distribution of~zeros for $\alpha = 1.5$.
Two nearly equally populated states at the~maximum of~the~envelope (here, the~ground state and the~first excited state) generate the~zeros closest to the~real-time axis.
Although the~approximate distribution places these zeros in~the~negative-$\beta$ half-plane, while the~exact zeros lie on the~opposite side and do not form a~perfectly periodic chain, it correctly captures the~large-scale zero density, in~particular the~presence of~zeros near the~real-time axis.
This is illustrated in~the~inset in~Fig.~\ref{fig:Models}(e), where the~exact and approximate distributions contain the~same number of~zeros in~the~displayed region.
The remaining differences are fine-structure effects beyond the~two-level approximation.
The role of~additional levels is discussed in~Appendix~\ref{app:deformation}, where we argue that their effect may actually decrease with increasing system size and improve the~predictions of~the~approximate distribution.

Fig.~\ref{fig:Models}(d) shows a~quench across the~QPT in~the~nearest-neighbor Ising model.
In this case, the~approximate zero distribution reproduces the~exact positions of the zeros, which can be obtained analytically~\cite{heyl2013dynamical}.
The red lines indicate the~exact zeros in~the~thermodynamic limit.
This perfect agreement follows from the~two-band form of~the~model and the BCS structure of the~initial state, as we explain in~the~next subsection.

\begin{figure}
    \includegraphics[width=0.99\linewidth]{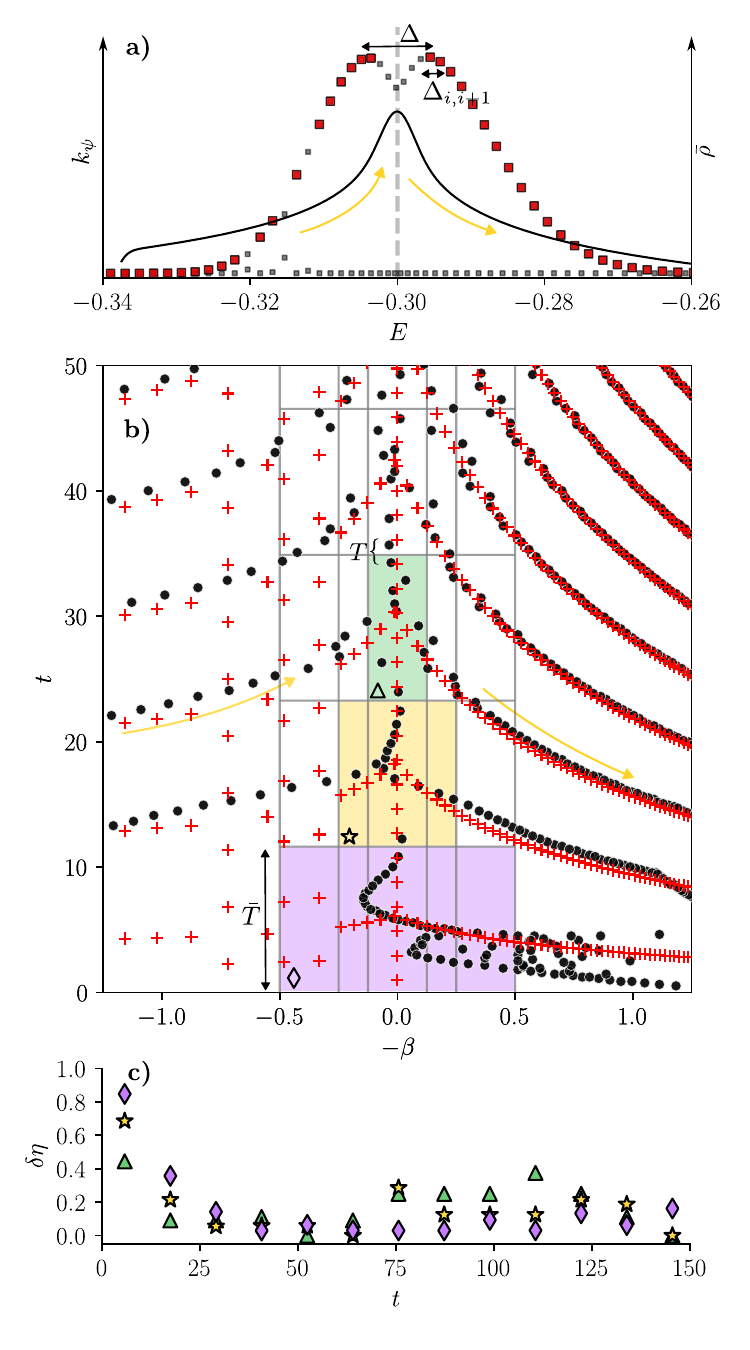}
    \caption{
    \textbf{(a)}~Energy distribution $k_{\psi}$ (gray squares) of~the~initial state in~the~basis of~the~final Hamiltonian $\hat{H}_f$ with the~envelope (red large squares).
    The~solid line shows the~smoothed level density $\bar{\rho}(E)$ of~the~final Hamiltonian.
    \textbf{(b)}~Exact (black dots) and approximate (red crosses) distributions of~zeros in~the~complex plane $(-\beta,t)$ of~the~survival amplitude of~the~initial ground state quenched from $h_i = 0.2$ to $h_f  = 0.6$ in~the~fully connected Ising model, $N = 400$. The~mean energy equals the~ESQPT energy, $E_{\text{ESQPT}} = -0.3$, marked by the~gray dashed line.
    \textbf{(c)}~Relative difference $\delta \eta$ between the~number of~exact and approximate zeros, evaluated in~rectangular boxes of~height $\bar{T}$ and different widths, shifted along the~real-time axis, as indicated in~panel (b) by matching colors and markers. The~time coordinate of~each point corresponds to the~center of~the~box. 
    }
    \label{fig:ESQPT}
\end{figure}

To demonstrate the~approximate zero distribution in~a~more complex setting, we 
consider a~quench to the~excited-state quantum phase transition (ESQPT) energy~\cite{cejnar2021excited} in~the~fully connected model.
In this model, the~ESQPT corresponds to the~energy at which the~constant-energy manifold in~classical phase space changes from disconnected to connected, reflecting the~underlying double-well structure of~the~Hamiltonian.

Quenching a~ground state from $h_i < h_f$ to  $h_f= \frac{h_i + 1}{2}$
produces a~bimodal energy distribution of~the~initial state~\cite{perez2011quantum}.
The dip in~the~distribution originates from the~classically forbidden region between the~wells.
The resulting energy distribution $k_{\psi}$ is plotted in~Fig.~\ref{fig:ESQPT}(a).
 
As shown in~Fig.~\ref{fig:ESQPT}(b), the~approximate distribution reproduces the~main features of~the~exact distribution, except at very short times.
In particular, it captures the~increased density of~zeros near the~real-time axis and the~branching lines of~zeros emerging from the~center.
According to the~dominance condition in~Eq.~(\ref{eq:inequality}), states lying in~the~dip of~the~energy distribution are excluded from the~envelope.
The two states at the~edges of~the~dip form the~maxima of~the~envelope and define an energy spacing $\Delta$, which generates a~dense chain of~zeros near the~real-time axis with period $T = \frac{2\pi}{\Delta}$.
Zeros generated by the~remaining terms in~the~envelope have larger periods derived from the~local level spacings $\Delta_{i,i+1}$, giving rise to the~branching structure. 
These zeros form lines that bend upward in~the~left half-plane and downward in~the~right half-plane.
This structure arises from the~energy dependence of~the~level spacings between the~states of~the~envelope, which decrease up to the~ESQPT energy and then increase across the~spectrum.
For this model, the~same trend is reflected in~the~level density of~the~final Hamiltonian, $\rho(E) = \sum_{j}\delta(E - E_j)$, as the~envelope is formed by almost all the~states in~the~given parity sector.
The connection between the~level density and the~structure of~the~branches is indicated by the~arrows in~the~plots.

A smoothed level density $\bar{\rho}$, obtained by replacing the~delta functions with Gaussians of~finite width, is shown in~Fig.~\ref{fig:ESQPT}(a) for comparison.

This example provides a~good test of~the~approximate construction, since many highly populated states in~the~dip of~the~energy distribution are excluded from the~envelope.
To quantify the accuracy of the approximation, we compare the numbers of exact and approximate zeros within rectangular boxes in the complex plane.
These boxes form a grid along the real-time axis as indicated in~Fig.~\ref{fig:ESQPT}(b).
For each box, we define the~relative difference
$
     \delta \eta = \frac{|\eta_e - \eta_a|}{\eta_e},
$
 where $\eta_e$ and $\eta_a$ denote the~number of~exact and approximate zeros in~that region.
 The~height of~these~boxes was chosen as $\bar{T} = \frac{2\pi}{\Delta_{i,i+1}}$, which is set by the~local spacing near the~maximum of~the~envelope.
 It approximately corresponds to the~period of~the~zeros in~the~branched lines rather than the~denser zeros near the~real-time axis, whose period $T$ is set by the~bimodal-like shape of~the~energy distribution.
 As shown in~Fig.~\ref{fig:ESQPT}(c), the~discrepancy is largest at short times, again due to the~nearly equidistant spectrum, and decreases at longer times.
 It is typically larger for smaller boxes, which are more sensitive to the~fine structure of~the~zero distribution.
 For box sizes comparable to the~short period $T$, the~comparison becomes sensitive to the~positions of~individual zeros rather than coarse-grained zero density, and $\delta\eta$ correspondingly exhibits strong oscillations and may even diverge.

\subsection{Two-band models: exact zero distribution from the~energy envelope}

The zero distribution of~the~survival amplitude of~a~quenched BCS ground state in~two-band models is well understood.
There, the~dynamics factorizes into independent two-level momentum subspaces.
Each subspace generates a~periodic chain of~zeros in~the~complex plane.

Here, we reinterpret this well-known mechanism in~terms of~the~energy envelope of~the~initial state.
This viewpoint explains why the envelope construction gives the exact distribution of zeros in two-band models.
By tracing the origin of the zero distribution to the envelope rather than to a factorization into independent two-level subspaces, it also explains the robustness of the distribution under perturbations and extends the same interpretation beyond noninteracting two-band models.

We consider the~energy distribution $k_{\psi}(E)$ of~the~initial ground state, $|\psi\rangle \equiv |\emptyset_i\rangle$, expressed in~the~eigenbasis of~the~final Hamiltonian $\hat{H}(h_f)$.
The eigenstates of~the~initial and final Hamiltonian are related by the~Bogoliubov transformation~\cite{10.21468/SciPostPhysLectNotes.82,schmitt2015dynamical}.
The overlap of~the~initial ground state with a~$2n$-particle eigenstate of~the~final Hamiltonian, populated by particle pairs with momenta in~the~allowed sector $\{\pm q_{j_1},\dots,\pm q_{j_n}\}$, can be written as 
\begin{equation}
    \label{eq:GS_overlap}
    \langle \emptyset_i|\{q_{j_1},\dots, q_{j_n}\}_f\rangle = \langle \emptyset_i|\emptyset_f\rangle \prod_{j} Z_{-q_j,q_j}.
\end{equation}
The factors $Z_{-q,q}$ are excitation amplitudes associated with the~creation of~quasiparticle pairs; see Appendix~\ref{app:twoband}.

The key observation is that the~envelope of~the~quenched ground state is formed by eigenstates of~the~final Hamiltonian with an increasing number of~pairs of~quasiparticles.
Neighboring states along the~envelope differ by the~occupation of~a~single quasiparticle pair with momenta $(-q,q)$.

The order of quasiparticle pairs entering the~envelope is determined by the~competition between their contribution to the~overlap and their associated energy cost, according to the~function
\begin{equation}
    \label{eq:Wfunction}
    W(q) = \frac{\ln |Z_{-q,q}|}{\varepsilon_q(h_f)},
\end{equation}
where $\varepsilon_q(h_f)$ is the~single-particle energy of~the~final Hamiltonian.
For each consecutive state in~the~envelope, the~additionally created pair yields the~largest overlap contribution per energy cost.

Exact expressions for excitation amplitudes and the explicit illustration for the~Ising model are given in~the~first section of~Appendix~\ref{app:twoband}.
There, we verify explicitly that the~states constructed according to the~ordering function~$W(q)$ satisfy the~envelope condition in~Eq.~(\ref{eq:inequality}) and therefore generate the~approximate zero distribution in~two-band models.

The derivation relies only on the~quasiparticle structure of~the~spectrum and does not depend on the~details of~the~Hamiltonian.
The resulting envelope construction therefore applies generally to noninteracting quasiparticle systems, where adding a~particle increases the~energy and changes the~overlap by a~factor, independently of~the~occupation of~other modes.

If the~envelope inequality holds strictly, two neighboring envelope states, which differ by a~pair of~quasiparticles with momenta $(-q,q)$, generate a~two-level chain of~zeros according to~Eq.~(\ref{eq:twilevelZero}).
Evaluating this expression using the~quasiparticle energies and overlaps of~a~BCS-type ground state yields
\begin{equation}
\label{eq:BCS_1D_zeros}
    z_n(q) = \frac{1}{\varepsilon_q(h_f)}\left [\ln |Z_{-q,q}| + i\pi \left(n+\frac{1}{2}\right)\right], n \in \mathbb{Z},
\end{equation}
which reproduces the~known distribution of~zeros of~quenched two-band models~\cite{schmitt2015dynamical} and connects the~envelope approach to the~topology interpretation of~DQPTs in~these models~\cite{PhysRevB.91.155127}.

\begin{figure}
    \centering
    \includegraphics[width=1\linewidth]{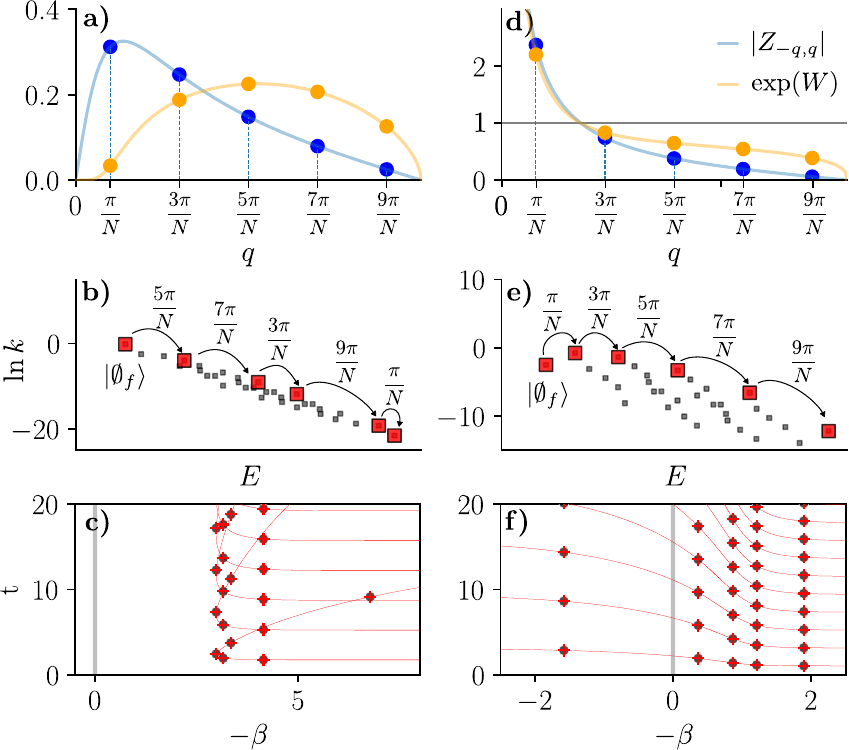}
\caption{The left column shows a~short quench of~the~ground state, $h_i = 0.1$ and $h_f = 0.2$, while the~right column shows a~long quench across the~QPT, $h_i = 0.1$ and $h_f = 0.5$, in~the~nearest-neighbor Ising model with $N = 10$.
\textbf{Top~panels:} Excitation amplitudes $|Z_{-q,q}|$ and the~ordering function $W(q)$, plotted as $e^{W(q)}$ to match the~common scale, as functions of~quasiparticle momentum. Dots mark the~allowed momentum values.
\textbf{Middle~panels:} Energy distribution (black squares) and envelope (red squares) of~the~initial states on a~logarithmic scale. Arrows indicate which quasiparticle pair distinguishes neighboring states along the~envelope.
\textbf{Bottom~panels:} Corresponding exact (black circles) and approximate (red crosses) zero distributions, together with the~lines of~zeros in~the~thermodynamic limit from~\cite{heyl2013dynamical}.}
    \label{fig:IsingMultiple}
\end{figure}

Fig.~\ref{fig:IsingMultiple} illustrates the~relationship between the~excitation amplitudes $Z_{-q,q}$, the~ordering function $W(q)$, and the~resulting envelopes for quenches within the~same phase and across the~QPT in~the~Ising model.
The shown example corresponds to a~system with $N = 10$ sites.
The symmetric ground state belongs to the~odd-momentum sector, where the~allowed momenta are odd multiples of~$\pi/N$.

The first state in~the~envelope is the~ground state of~the~final Hamiltonian.
Consecutive states along the~envelope are obtained by adding quasiparticle pairs, as illustrated by the~arrows in~Fig.~\ref{fig:IsingMultiple}(b) and Fig.~\ref{fig:IsingMultiple}(e), with momenta $(-q,q)$ in~the~order prescribed by the~ordering function~$W$ in~Fig.~\ref{fig:IsingMultiple}(a) and Fig.~\ref{fig:IsingMultiple}(d).

For a~short quench, the~magnitude of~the~excitation amplitude~$|Z_{-q,q}|$ is below one.
Adding a~quasiparticle pair therefore always decreases the~population of~subsequent envelope states, resulting in~a~monotonically decreasing envelope.
As a~consequence, the~zeros are distributed only in~the~negative-$\beta$ half-plane.

By contrast, for a~quench across the~QPT, the~excitation amplitude~$|Z_{-q,q}|$ crosses the~value 1. 
Creating quasiparticle pairs with $|Z_{-q,q}| > 1$ increases the~population of~successive envelope states and produces an increasing segment of~the~envelope.
This increasing part generates zeros in~the~positive-$\beta$ half-plane.

For a~BCS ground state in~a~two-band model, the~survival amplitude factorizes into independent two-level momentum subspaces
\begin{equation}
\label{eq:twobandBCSsurvivalamplitude}
\mathcal{L}(z) 
= \prod_{\{q\}} \langle \emptyset_i(q)|e^{-z\hat{H}_f(q)}|\emptyset_i(q)\rangle \equiv \prod_{\{q\}} \mathcal{L}_q(z).
\end{equation} 
The full distribution of~zeros is therefore the~union of~the~two-level zeros generated by each momentum mode and provides a~limiting case in~which the~approximate zero distribution derived from the~envelope becomes exact.

\subsection{Energy envelope in~the~collective regime and finite-size effects}

From the~envelope perspective, zeros approach the~real-time axis when the~maximum of~the~energy distribution is formed by two or more equally populated eigenstates.
For ground-state quenches, the~distribution is initially maximal at the~final ground state.
As the~quench amplitude increases, the~maximum shifts as a~low-lying excited state becomes equally populated with the~ground state.
Under this condition, the first zeros approach the~real-time axis.

We apply this principle to explain the~behavior of~the~zero distribution in~the~fully connected Ising model.
In this model, the~zeros appear on the~real-time axis in~the~thermodynamic limit even for arbitrarily short quenches.
The dynamics are dominated by the~collective degree of~freedom.
In the~classical limit, which coincides with the~infinite-size limit, the~system is described by a~classical Hamiltonian.
If the~initial state is localized around the~global quadratic minimum, it is well approximated by a~coherent state.
Its energy distribution in~the~eigenbasis of~the~post-quench Hamiltonian is approximately Poissonian.
This approximation can fail in~the~presence of~nontrivial phase-space structures, for instance, for quenches near an ESQPT, as shown already in~Fig.~\ref{fig:ESQPT}.

A key feature of~a~Poisson distribution is that its maximum lies close to its mean.
Although the~maximum of~the~post-quench energy distribution is generally difficult to determine directly, the~mean energy is easily obtained.
For a~sudden quench $h_i \rightarrow h_f$ in~a~Hamiltonian of~the~form $\hat{H}(h) = \hat{H}_0 + h \hat{V}$, 
the mean energy of~the~initial state after the~quench shifts linearly with $\Delta h = h_f- h_i$
\begin{equation}
\label{eq:MeanEnergy}
\langle \hat{H}(h_f) \rangle = \langle \hat{H}(h_i) \rangle + \Delta h \langle \hat{V}\rangle.
\end{equation}
For further discussions, see Refs.~\cite{kloc2018quantum,PhysRevA.103.032213,PhysRevA.104.053722}.
Whenever the~post-quench distribution remains close to Poissonian, its maximum follows the mean closely.
This provides a~simple way to estimate how the~quench amplitude shifts the~maximum of~the~distribution.

To compare this estimate with the~actual maximum, we introduce a~simple diagnostic.
For a~ground-state quench, an approximate chain of~zeros first appears when a~higher excited state $|E_j^{(f)}\rangle$ becomes equally populated with the~ground state $|E_0^{(f)}\rangle$ of~the~final Hamiltonian, that is, when $k_0(h_f) = k_j(h_f).$
We monitor this condition through the~ratio
 \begin{equation}
     \label{eq:ratio}
     R = \frac{k_{\text{max}_2}}{k_{\text{max}_1}},
 \end{equation}
 where $k_{\text{max}_1}$ and $k_{\text{max}_2}$ are the~largest and second-largest populations in~the~envelope, respectively.
 The~condition $R = 1$ signals the~presence of~approximate zeros on the~real-time axis.

 \begin{figure}
    \includegraphics[width=0.95\linewidth]{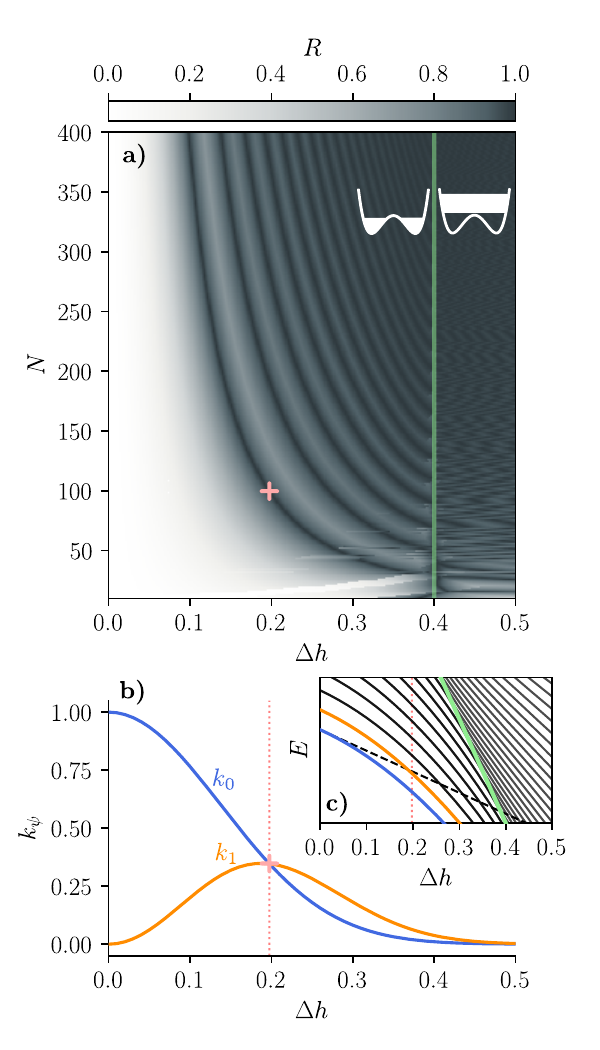}

    \caption{\textbf{(a)} Ratio $R$ as a~function of~the~quench length $\Delta h$ and system size $N$ for a~symmetric ground state at $h_i=0.2$.
    The~green line marks $\Delta h=0.4$, where the~initial state is quenched to the~ESQPT energy.
    Quenches on the~two sides of~this line end in~regimes with disconnected and connected constant-energy manifolds, respectively, as indicated by the~pictograms.
    \textbf{(b)} Populations of~the~ground state $k_0$ and first excited state $k_1$ as functions of~$\Delta h$ for $N=100$.
    The~dotted line marks their intersection at $\Delta h^*\approx0.198$, also indicated by a~pink cross in~panel (a).
    \textbf{(c)} Energy levels of~$\hat{H}_f$ versus $\Delta h$, with the~ground state and first excited state highlighted.
    The~dashed line shows the~mean energy of~the~quenched state from Eq.~(\ref{eq:MeanEnergy}), the~vertical dotted line marks $\Delta h^*$, and the~angled line indicates the~ESQPT energy.}
    \label{fig:Ns}
\end{figure}

For illustration, we consider a~quench in~the~fully connected Ising model that remains within the~same ground-state phase, i.e., the~quench does not cross the~ground-state QPT at $h = 1$.
Fig.~\ref{fig:Ns}(a) shows the~ratio $R$ as a~function of~the~length of~the~quench $\Delta h$ and the~system size $N$. 
The quench length required to satisfy the~condition $R=1$ decreases with increasing system size and vanishes in~the~infinite-size limit \cite{anomalous2017}.
 The~apparent lines $R = 1$ in~Fig.~\ref{fig:Ns}(a) are a~consequence of~the~discreteness of~the~envelope, as each line corresponds to a~different pair of~equally populated consecutive levels in~the~envelope. 

Panels (b) and (c) in~Fig.~\ref{fig:Ns} illustrate this behavior for $N = 100$.
The populations of~the~ground state and the~first excited state of~the~final Hamiltonian become equal at $\Delta h \approx 0.198$, signaling the~formation of~an approximate chain of~zeros on the~real-time axis.
This occurs near the~crossing point of the~mean energy of~the~quenched state and the~first excited level, consistent with the~maximum following the~mean of~the~distribution along the~quench.

For small system sizes ($N \sim 20$), the~condition $R = 1$ is satisfied only for quenches across the~ESQPT energy, here $\Delta h = 0.4$.
In this regime, the~limited energy resolution of~the~finite system implies that higher-energy states are efficiently excited only when the~quench takes the~system into a~different excited-state phase with a~distinct topology of~manifolds of~constant energy in~the~phase space (as depicted by the~sketch in~Fig.~\ref{fig:Ns}(a)).

As the~system size increases, this finite-size effect is gradually washed out.
Within the~present framework, this indicates that the~appearance of~zeros on the~real-time axis is not generically tied to crossing a~ground-state or excited-state phase transition.
Instead, it follows from the~shift of~the~maximum of~the~energy distribution induced by the~quench.
For ground-state quenches in~collective models, this shift occurs straightforwardly, since the~energy distribution develops a~Poissonian-like shape with a~mean moving linearly with the~quench length.

\section{Short-Time zero distributions and Anomalous DQPTs}
\label{sec:IV}
In this section, we analyze the~zero distribution of~the~survival amplitude $\mathcal{L}$ in~the~short-time regime.
At short times, the~survival amplitude in~Eq.~(\ref{def:overlapComplex}) is governed by a~coherent sum of~the~contributing terms.
This gives rise, in~general, to a~distribution of~zeros that differs from the~approximate one formed by cancellation of~the~two or few largest terms.
As time increases, dephasing induced by the~generally non-equidistant energy spectrum progressively destroys phase coherence. Consequently, the~window of~energy levels with correlated phases shrinks, and the~distribution of~zeros continuously evolves toward its asymptotic form that resembles the~approximate distribution.

To capture the~close-to-coherent dynamics, we model the~survival amplitude using an almost equidistant spectrum together with a~Gaussian energy distribution of~the~initial state.
We show that this minimal model reproduces the~characteristic features associated with ADQPT.

In the~case of~an equidistant spectrum with level spacing $\Delta$, the~modulus of~the~survival amplitude is periodic in~time.
To model the~effect of~dephasing on the~evolution of~the~distribution of~zeros at short times, we break the~uniformity by a~small perturbation $\varepsilon$ and consider a~spectrum with levels
\begin{equation}
    E_j = E_{\text{GS}} + j\Delta  + j^2 \frac{\varepsilon}{2}, \quad j\in \{j_i,\dots,j_f\},
\end{equation}
with slowly varying level spacing $\Delta_j = \Delta + \varepsilon (j+\frac{1}{2})$.
We focus on the~regime $\Delta \gg \varepsilon$, 
such that the~level spacing remains positive over the~relevant energy window.
The parameter $E_{\text{GS}}$ sets the~energy of~the~ground state and we fix it to $E_{\text{GS}} = 0$.

We further assume an initial state with a~Gaussian energy distribution 
\begin{equation}
    k_j  \propto e^{-\frac{(j-\mu)^2}{2\sigma^2}},
\end{equation}
where the parameter $\mu$ is introduced to improve potential further fit to a~given physical spectrum, but for the~analytical treatment we set $\mu = 0$.
This setting provides a~minimal model for the~survival amplitude \cite{Lerma-Hernandez_2018} of~a~finite $(d+1)$-level system
\begin{equation}
    \label{eq:GaussianModelSurAmplitude}
        \mathcal{L}_G(z) = \frac{1}{\mathcal{N}} \sum_{j=j_i}^{j_f = j_i+d}  e^{ -\frac{j ^2}{2\sigma^2(z)}}  e^{-z\Delta j}, 
\end{equation}
where the~coefficients are Gaussian distributed with complex-time-dependent width 
$\sigma^2(z) = \frac{\sigma^2}{1 + z \varepsilon  \sigma^2}$ and $\mathcal{N}$ is a~normalization constant.

The survival amplitude $\mathcal{L}_G$ of~the~Gaussian model at short times is governed by a~coherent sum over many nearby levels and exhibits two distinct regimes of~zeros.
The first corresponds to an initial state centered deep in~the~interior of~the~spectrum, where the~boundedness imposed by the~ground state or the~most excited state can be neglected.

\begin{figure}
    \centering
    \includegraphics[width=0.94\linewidth]{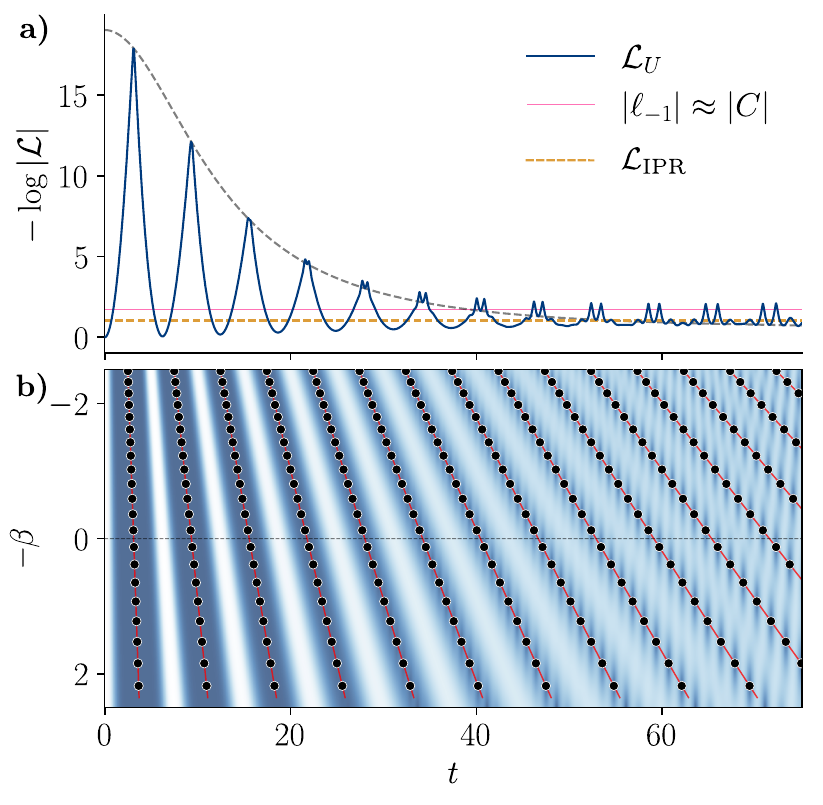}
    \caption{\textbf{(a)}~The survival amplitude of~the~unbounded Gaussian model at $\beta = 0$, plotted on a~$-\log$ scale to enhance the~behavior near minima.
    The~dashed curve shows the~decay governed by $\bar\Theta_{\mathrm{Decay}}$ as the~survival amplitude approaches the~IPR value (horizontal dashed line).
    The~thin horizontal line corresponds to $-\log|\ell_{-1}|$ and approximates the~cutoff term $C$. Their crossing indicates the~time beyond which $\mathcal{L}_U$ becomes dominant over the~cutoff term $C$ in~the~bounded model.
    \textbf{(b)}~Corresponding zero distribution (black dots) of~$\mathcal{L}_U$ in~the~complex plane.
    Red lines indicate the~analytic curves given by Eq.~(\ref{eq:Model_Lines}).}
    \label{fig:unboundGaussian}
\end{figure}

In this regime, the~distribution of~zeros can be determined analytically, as shown in~Appendix \ref{sec:ShortTimes}.
Remarkably, it is given by the~two-level formula in~Eq.~(\ref{eq:twilevelZero}),
as each pair of~consecutive levels $E_j$ and $E_{j+1}$ gives rise to a~chain of~zeros, given by 
\begin{equation}
    \label{eq:ShortTimeModel_zeros}
    z_{j,j+1}(n) = \frac{1}{\Delta_j}\left[ -\frac{2j - 1}{2\sigma^2}  +  2\pi i \left(n+\frac{1}{2}\right)\right], \quad n \in \mathbb{Z},
\end{equation}
along the~real-time axis (see Fig.~\ref{fig:unboundGaussian}).
This behavior follows from the~symmetry of~the~Gaussian energy distribution about its maximum.

As a~result, exact and approximate zeros coincide in~this model on all time scales.
This is consistent with the~behavior observed in~Fig.~\ref{fig:Models}(a) for the~long quench in~the~fully connected Ising model.

In the~second regime, the~center of~the~Gaussian distribution is close to the~ground state.
The energy distribution of~the~initial state is highly asymmetric, and the~corresponding zeros no longer form the~two-level chains of~Eq.~(\ref{eq:ShortTimeModel_zeros}).
Consequently, the~distribution of~zeros at short times deviates from the~approximate one, as seen in~Fig.~\ref{fig:Models}(b), but gradually evolves toward it at later times.

To explain the gradual evolution towards the~two-level distribution, we express the~survival amplitude of~the~bounded Gaussian model as a~truncated sum of~an auxiliary unbounded Gaussian model with survival amplitude $\mathcal{L}_U = \sum_{-\infty}^{\infty} \ell_{j}$ with zeros given by Eq.~(\ref{eq:ShortTimeModel_zeros}).
The survival amplitude of~the~bounded model $\mathcal{L}_{G}$ can then be written in~the~form
\begin{equation}
    \label{eq:bound_model}
    \mathcal{L}_G(z) \propto \mathcal{L}_U(z) - C(z),
\end{equation}
where 
\begin{equation}
    C(z) = \sum_{j<j_i}\ell_j(z) + \sum_{j>j_f}\ell_j(z)
\end{equation}
collects the~cutoff contributions that restore the~boundedness of~the~original spectrum.

It is precisely the~cutoff term $C$ that causes the~short-time zero distribution to deviate from the~two-level zeros of~the~unbounded amplitude $\mathcal{L}_U$.
The cutoff term dominates over $\mathcal{L}_U$, $|C| > |\mathcal{L}_U|$, in~the~neighborhood of~the~periodic zeros of~$\mathcal{L}_U$.
This is illustrated in~Fig.~\ref{fig:unboundGaussian}(a), where both terms are shown on a~logarithmic scale and the~cutoff term is approximated by its maximal value for clarity.
Hence, in~this bounded regime, the~zeros of~the~survival amplitude $\mathcal{L}_G$ arise from the~mutual cancellation of~the~two terms in~Eq.~(\ref{eq:bound_model}), rather than being deformed zeros of~one of~the~terms.

However, this changes over time due to dephasing.
The magnitude of~$\mathcal{L}_U$ grows in~the~vicinity of~its zeros as the~survival amplitude approaches the~asymptotic value given by the~inverse participation ratio (IPR)
\begin{equation}
\label{eq:IPR}
    \mathcal{L}_{\text{IPR}} = \sqrt{\sum_{j=j_i}^{j_f} k_j^2}.
\end{equation}
The loss of~coherence due to the~nonuniform level spacing, which can be understood as a~shrinking of~the~effective width $\sigma(z)$, causes the~decay of~the~peaks around the~zeros of~$\mathcal{L}_U$, as shown in~Fig.~\ref{fig:unboundGaussian}(a). 
The decay is captured by the~$\bar{\Theta}_{\text{Decay}}$ function, which we derive in~Appendix~\ref{sec:ShortTimes}.
As a~result, $\mathcal{L}_U$ gradually dominates over $C$ in~the~vicinity of~its zeros.
The zeros of~the~survival amplitude $\mathcal{L}_G$ are then given by those of~$\mathcal{L}_U$, perturbed by the~cutoff term $C$.
This crossover can be understood as a~deformation of~the~initial distribution of~zeros into the~long-time distribution, in~the~sense of~Rouché’s theorem.

\begin{figure}
    \centering
    \includegraphics[width=0.95\linewidth]{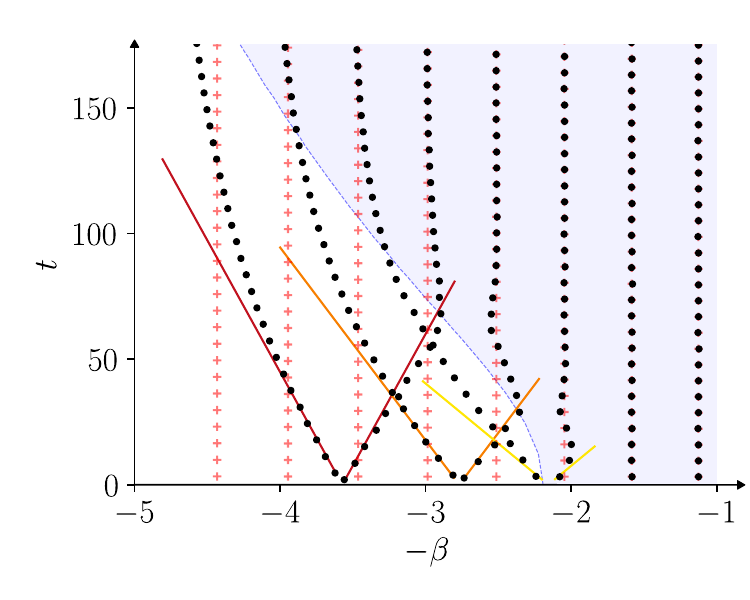}
    \caption{Zero distribution (black dots) of~the~survival amplitude of~the~Gaussian model with parameters $\varepsilon = 5\cdot10^{-3}$, $\Delta = 1$ and $\sigma = 1.5$.
     The~lines show the~trajectories of~the~zeros to first order in~$\varepsilon$.
    The~initial conditions $z_0$ correspond to those in~Fig.~\ref{fig:boundGaussian}(a) in~Appendix~\ref{sec:ShortTimes} marked by corresponding colors. 
    The~red crosses show the~zero distribution of~the~unbounded survival amplitude $\mathcal{L}_U$ with the~same parameters.
    The~blue area marks the~region where the~zeros of~the~bounded survival amplitude can be understood as deformed zeros (red crosses) of~the~unbounded model, i.e. $\bar{\Theta}_{\text{Decay}} > |\ell_{-1}|$. 
    }
    \label{fig:runningZeros}
\end{figure}

Fig.~\ref{fig:runningZeros} illustrates this crossover with two distinct regions highlighted.
In the~blue-shaded region, the~cutoff term only weakly perturbs the~two-level zeros of~$\mathcal{L}_U$; in~the~white region, the~boundedness of~the~spectrum determines the~positions of the~zeros directly.
As time increases, the~cutoff-dominated white region shrinks, and the~zero distribution gradually approaches the~asymptotic two-level form.

This deformation can also be understood as a~dephasing effect.
The nonuniform spectrum perturbs otherwise periodic chains of~zeros.
As dephasing accumulates from one period to the~next, the~perturbation changes slightly, shifting the~zeros away from their periodic positions and causing them to trace curved trajectories in~the~complex plane rather than aligning along a~line of~constant $\beta$.

To leading order in~$\varepsilon$, the~trajectory of~a~zero 
$$z(n,\varepsilon) = z_0(n) + \varepsilon\delta z(n) + \mathcal{O}(\varepsilon^2),$$
 corresponds to a~line, as shown in~Fig.~\ref{fig:runningZeros}.
Here, $z_0(n)$ denotes the~zeros of~the~periodic survival amplitude, with each time period labeled by the~integer $n$.
For $\varepsilon \neq 0$, each period introduces a~small displacement $\delta z(n)$.
The exact form is derived in~Appendix~\ref{sec:ShortTimes}.
The zero $z_0(n = 0)$ acts as the~initial condition for the~whole family of~discretely evolving zeros.

\begin{figure}
    \centering
    \includegraphics[width=\linewidth]{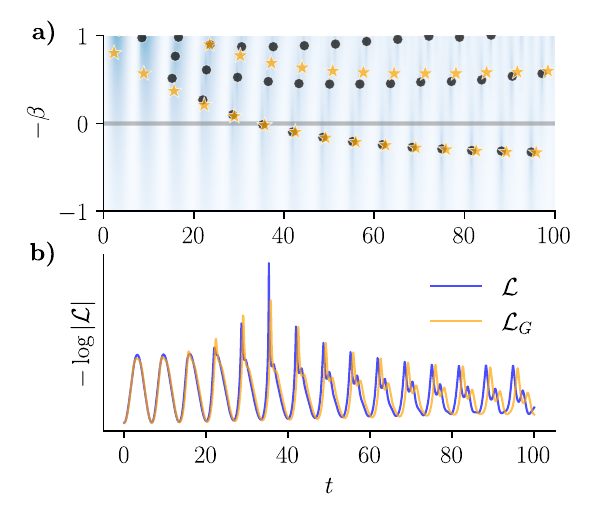}
    \caption{\textbf{(a)} Distribution of~zeros for a~ground-state quench in~the~fully connected Ising model (black dots) with $h_i = 0$ and $h_f = 0.2$ for $N = 100$, compared with the corresponding fitted bounded Gaussian model $\mathcal{L}_G$ (orange stars).
    \textbf{(b)} Comparison of the survival amplitudes of the fully connected Ising model and the fitted Gaussian model on a $-\log$ scale.
    The fitted spectral parameters are $\Delta = 0.958$ and $\varepsilon = -0.022$; the fitted energy-distribution parameters are $\sigma = 1.3$ and $\mu = 0.5$, with $j \in \{0,1,\dots,15\}$.
    }
    \label{fig:LipkinRunning}
\end{figure}

The trajectory picture developed above provides a~natural interpretation of~the~ADQPT.
In systems where dephasing is sufficiently slow, the~zeros of~the~survival amplitude do not appear at their asymptotic positions immediately, but instead approach the~real-time axis along these trajectories.
In the~following, we demonstrate this mechanism explicitly for the~fully connected Ising model, where ADQPTs were originally reported.

A comparison of~the~survival amplitude and distributions of~zeros for a~ground-state quench in~the~fully connected Ising model and in~the~corresponding bounded Gaussian model is shown in~Fig.~\ref{fig:LipkinRunning}.
The parameters of~the~Gaussian model---the width $\sigma$, the~shift $\mu$ at $\beta = 0$, and the~effective level spacing $\Delta$ 
 with weak non-uniformity $\varepsilon$---are obtained by fitting the~energy distribution of~the~initial state.

The bounded Gaussian model reproduces the~zero distribution of~the~fully connected Ising model near the~real-time axis,
including the~delayed arrival of~the~first zero, which is the~defining feature of~the~ADQPT.
This agreement demonstrates that the~evolution of~the~short-time distribution of~zeros is controlled by slow dephasing in~a~nearly equidistant, bounded spectrum, rather than by microscopic details of~the~model.
In the~present case, the~asymmetry of~the~energy distribution originates from a~short quench that prepares the~initial state close to the~ground state of~the~final Hamiltonian.

As shown in~Ref. \cite{anomalous2017}, for ground-state quenches the~range of~quench amplitudes over which the~distribution of~zeros exhibits signatures of~the~ADQPT shrinks as the~interaction range in~the~Ising model is reduced toward the~nearest-neighbor limit, and eventually disappears.

The absence of~late-arriving zeros in~generic short quenches---where the~resulting distribution of~zeros remains highly asymmetric, as in~Fig.~\ref{fig:Models}(c)---can be understood from the~relation between the~dephasing time scale and the~time scale for the~arrival of~the~first zero.
In the~Gaussian model, the~dephasing time scale is controlled by the~parameter $\varepsilon$.
When the~populated levels in~the~envelope are far from equidistant (corresponding to large $\varepsilon$), dephasing becomes sufficiently fast that the~deformation of~the~short-time distribution into the~asymptotic two-level structure occurs on time scales shorter than the~arrival of~the~first zero, $t \sim \Delta^{-1}$.

\section{Summary and Conclusions}\label{sec:V}

We have identified the~envelope of~the~conserved initial-state energy distribution as the~governing structure of~the~zeros of~the~complex-time survival amplitude in~a~generic finite closed quantum system.
We have established this connection by constructing an approximate zero distribution of~the~survival amplitude based solely on this envelope.
The approximation is built from chains of~periodic two-level zeros associated with the~dominant contributions to the~survival amplitude and reproduces the~large-scale density and qualitative structure of~the~exact distribution of~zeros in~the~complex plane.
This construction shows that the~energy envelope governs the~global distribution of~zeros.

In this framework, the~approximate zeros reach the~real-time axis once the~maximum of~the~energy distribution is formed by two or more equally populated states.
This condition has been observed in~various models~\cite{PhysRevB.106.045410,Cao_2024}, supporting the~view that finite-size precursors of~DQPTs are governed by the~shape and deformation of~the~energy distribution.
The envelope approach thus connects the~shape of~the~energy distribution directly to the~emergence of~dynamical quantum phase transitions and provides a~natural interpretation of~this effect.

We demonstrated the~generality of~this interpretation in~the~1D Ising model with tunable interaction range.
There, the~approximate zero distribution qualitatively reproduces, across all interaction ranges, the~exact zero distributions of~the~survival amplitude for initial states prepared by sudden quenches in~the~transverse field.
In addition, we proved that the~approximate distribution becomes exact for ground-state quenches of~BCS states in~two-band noninteracting models.
In this way, the~envelope captures the~effect of~crossing the~phase boundary through the~deformation induced in~the~energy distribution along the~quench.

In the~collective regime, the~initial state localized near the~global minimum is well approximated by a~coherent state, and its energy distribution therefore closely follows a~Poisson distribution.
In this case, the~condition for zeros to approach the~real-time axis is approximately equivalent to the~mean energy of~the~quenched state crossing the~low-lying excited levels.
This is why the~appearance of~DQPTs in~such systems is generally not tied to ground-state or excited-state quantum phase transitions.
Instead, the~relevant quantity is the~deformation of~the~energy envelope induced by the~quench.

The envelope construction has clear limitations.
In systems with nearly equidistant spectra, the~short-time zero pattern can deviate strongly from the~asymptotic two-level structure because phase coherence persists at early times.
Using a~minimal Gaussian model, we showed that the~combination of~a~nearly equidistant spectrum and a~strongly asymmetric energy distribution, naturally produced by short ground-state quenches, generates an initial zero pattern that relaxes only gradually toward the~approximate distribution due to dephasing.
This provides a~simple mechanism for anomalous DQPTs as a~delayed approach of~zeros to the~real-time axis.

More generally, because the~approximate distribution is built from chains of~periodic two-level zeros, it captures the~qualitative structure and density of~zeros.
However, it does not predict their exact positions, in~particular, whether a~zero lies exactly on the~real-time axis for a~given finite system size.
In this sense, the envelope provides an organizing structure of finite-size precursors of DQPTs, and this correspondence becomes exact in noninteracting two-band models, even for a continuous spectrum.
Extending this understanding to the thermodynamic limit of generic interacting systems still requires further analysis.

\section{Acknowledgments}
J.N. thanks A. Polkovnikov and S. Kehrein for valuable discussions.
J.N. and J.S. acknowledge support from the~Charles University Grant Agency (GA UK
no.~215323).
P.S. and P.C. acknowledge support from the~Czech Science Foundation under Project No.~25-16056S.

\section*{Data availability}
The code used to produce the~exact and approximate zero distributions presented in~this work is publicly available at \cite{jan_strelecek_2026_19734887}. No additional datasets were generated.

\appendix

\section{Numerical calculation of~exact zeros}\label{app:numerics}
We briefly describe the~numerical method used to locate the~zeros of~the~survival amplitude $\mathcal L(z)$ in~the~complex plane.

Because of~the~complicated oscillatory behavior of~$\mathcal L(z)$, searching for zeros using common optimization techniques is numerically unstable. Instead, we exploit the~principle that for any simply connected 
region $R$ with boundary $\partial R$, the~number of~zeros of~
$\mathcal{L}(z)$ inside $R$ is given by the~winding number
\begin{equation}
    W_R = \frac{1}{2\pi i} \oint_{\partial R} 
    \frac{\mathcal{L}'(z)}{\mathcal{L}(z)} \, dz.
\end{equation}
Since $\mathcal{L}(z)$ is a~finite sum of~
exponentials, it has no poles, and $W_R$ counts exactly the~number 
of zeros inside $R$, including their multiplicity.

A simple approach for locating zeros is to compute $W_{R_i}$ around each square in~a~sufficiently dense grid and return those yielding a~nonzero value. This algorithm, however, is wasteful, since most squares contain no zeros. We instead explore the~complex plane iteratively.

We locate the~zeros using the~following recursive subdivision algorithm:\begin{enumerate}
    \item Partition the~region of~interest $R$ into a~$k \times k$ grid 
    of~subrectangles $\{R_1, \dots, R_{k^2}\}$.
    \item Evaluate $W_{R_i}$ for each subrectangle and discard all 
    cells with $|W_{R_i}| \leq \epsilon$, to account for a~numerical error threshold $\epsilon>0$.
    For the~calculations, we used the~numerical threshold $\epsilon = 0.2$.
    \item Take each remaining subrectangle as a~new region of~interest and repeat step~1.
    \item When sufficient precision is reached, terminate the~recursion and return the~centers of~the~final subrectangles, which now approximate the~locations of~zeros.
\end{enumerate}

\section{Supporting analysis for the~approximate distribution of~zeros}\label{app:deformation}
In this appendix, we provide technical details and heuristic arguments supporting the~construction of~the~approximate zero distribution presented in~Sec. \ref{sec:II}.

\subsection{Density of~zeros}
In this section, we show that the~exact distribution of~zeros inherits its coarse-grained density on largest scales from the~distribution generated by the~edge terms of~the~spectrum.
A key property of~the~approximate zero distribution constructed from two-level chains is that it reproduces this density on large scales in~the~complex plane.

Given an initial state $|\psi\rangle$, let $E_0$ and $E_d$ be lowest- and highest-energy levels populated by this state, with occupations $k_0, k_d > 0$.
We refer to these as the~edge states of~the~initial-state energy distribution.

Then, by Rouché's theorem, the~zeros of~the~survival amplitude of~the~initial state can be deformed onto the~zeros of~the~survival amplitude of~a~two-level system formed by the~edge terms.
\begin{equation}
\label{def:edgeSA}
\mathcal{L}_{\mathrm{edge}}(z) = \ell_0(z) +\ell_d(z)
\end{equation}
as the~condition $|\mathcal{L}_{\mathrm{edge}}| > |\mathcal{L}|$ is satisfied outside a~strip in~the~complex plane bounded by $\beta_0$ and $\beta_d$ defined as:
\begin{equation}
    r_0(\beta_0) = \sum_{j>0}r_j(\beta_0), \quad
r_d(\beta_d) = \sum_{j<d}r_j(\beta_d).
\end{equation}
As a~consequence, the~exact zeros of~the~survival amplitude of~the~initial state $|\psi\rangle$ are confined to this strip.

The zero distribution of~this edge system is given by Eq.~(\ref{eq:twilevelZero}) with zeros located at $\beta_{0d}$ and period $T_{0d}$, which is inversely proportional to the~size of~the~full spectrum.

Although the~total number of~zeros in~the~complex plane is infinite, their distribution can be characterized by the~density of~zeros in~time, defined as the~number of~zeros in~a~strip $\beta \in (-\infty, \infty)$ over a~time interval $T$. 
For the~edge two-level distribution, there is one zero over each time period $T_{0d}$.
On time scales larger than the~inverse mean level spacing, the~exact distribution of~zeros exhibits approximately the~same density.

The approximate zero distribution constructed from two-level chains generated by the~energy envelope reproduces this density.
To see this, we assume the~following partition of~the~spectrum by the~terms in~the~envelope:
$\Delta_{0d} = \sum_j \Delta_{j,j+1}$, where $\Delta_{j,j+1}$ 
are the~spacings between consecutive levels in~the~envelope.
The number of~zeros associated with the~chain generated by levels $E_j$ and $E_{j+1}$
up to time $t$ is approximately 
$$n_j(t) \approx t T_{j,j+1}^{-1} 
= |\Delta_{j,j+1}| \frac{t}{2\pi},$$
where the~exact number is, of~course, an integer.
Summing over all chains yields
\begin{equation}
\label{eq:approxdensity}
   \sum_j n_j(t) \approx
   \sum_j \frac{t|\Delta_{j,j+1}|}{2\pi} =  \frac{t|\Delta_{0d}|}{2\pi} \approx n_{\text{edge}}(t),  
\end{equation}
which shows that the~approximate distribution can be interpreted as a~redistribution of~the~edge-system zeros among the~chains generated by neighboring envelope terms.

This construction assumes that the~initially populated energy levels are not equidistant or nearly equidistant, so that the~survival amplitude is generically a~noncoherent sum of~exponentials and its local reduction to dominant terms is justified.

In the~equidistant case, the~distribution of~zeros differs qualitatively from the~lattice-like approximate distribution constructed from two-level chains.
The survival amplitude then effectively reduces to a~Fourier series of~the~energy distribution, leading to a~distinct zero structure.
Nevertheless, the~principle that the~exact distribution of~zeros shares the~same zero density as the~edge two-level system still holds.
For a~spectrum of~$d+1$ equidistant levels, $\Delta_{0d} = d\Delta$, both the~exact and the~edge distributions contain $d$ zeros within the~strip $[\beta_0,\beta_d]$ over one period $2\pi/\Delta$.

\subsection{Multilevel distribution of~zeros}
Here we discuss in~more detail the~distribution of~zeros connected with multilevel intersections of~the~terms from the~envelope.

When $m+1$ terms in~the~survival amplitude satisfy the~envelope inequality Eq.~(\ref{eq:inequality}) simultaneously in~a~strict sense, we can rewrite their sum as 
\begin{equation}
 \label{eq:multiple}
     \sum_{i = 0}^m \ell_{j_{i}}(z) = r_{j_0}(\beta_{j_{0}j_{m}})\sum_{i = 0}^m e^{-E_{j_i} (\beta' + i t)},
 \end{equation}
where $\beta' =\beta + \beta_{j_{0}j_{m}}$.
Apart from the~overall prefactor, the~approximate survival amplitude of~this reduced effective multilevel system therefore has the~form of~a~canonical partition function.
This representation allows one to interpret the~multilevel intersection as a~set of~terms that become equally populated at $\beta_{j_{0}j_{m}}$, i.e., 
 $r_j(\beta_{j_{0}j_{m}}) = r_{j_0}(\beta_{j_{0}j_{m}})$ for all the~terms forming the~given multilevel intersection.

The formation of~a~multilevel intersection is directly tied to the~shape of~the~envelope.
It occurs when the~populations of~the~participating consecutive levels depend exponentially on energy.
For a~set of~consecutive levels that forms the~multilevel intersection,
their populations $k_j$ depend on energy as
\begin{equation}
\label{def:multilevelinequality}
k_{j} = k_{j_0} \left( \frac{k_{j_0}}{k_{j_m}} \right)^{-\frac{\Delta_{j_0 j}}{\Delta_{j_0 j_m}}} \equiv \kappa_{0} \,\kappa^{\frac{E_{j}}{E_{j_0} - E_{j_m}}}.
\end{equation}
See the~sketch in~Fig.~\ref{fig:Theory}(b).
A special case is $\kappa = 1$, which corresponds to several equally populated states at the~maximum of~the~envelope, producing a~flat maximum and an approximate multilevel zero at the~real-time axis.

Although the~zeros of~the~expression in~Eq.~(\ref{eq:multiple}) cannot be obtained analytically, they can be viewed as deformed two-level zeros generated by the~edge terms $\ell_{j_0}$ and $\ell_{j_m}$.
Therefore, multilevel zeros are approximately located around $\beta_{j_0j_m}$ with an approximate period $T_{j_0j_m}$ given by Eq.~(\ref{def:betaANDperiod}).

A particularly simple case occurs when the~$m+1$ participating levels are equally spaced in~energy with spacing $\Delta$.
Equation (\ref{eq:multiple}) then reduces to a~geometric sum, yielding zeros at 
\begin{equation}
\label{eq:geometric}
z(n,n_0)
= \beta_{j_0j_m} + \frac{2\pi}{\Delta} \left( \frac{n}{m} + \frac{n_0}{m+1} \right),   
\end{equation}
where $n \in \mathbb{Z}$ labels successive periods and $n_0 \in \{1,\dots m\}$ labels the~zeros within one period.
This characteristic pattern of~zeros may be relevant for scarred initial states~\cite{PhysRevResearch.5.033090}, where the~envelope is expected to be formed by multiple nearly uniformly spaced states.

\subsection{Role of~additional terms}
At first sight, the~approximation underlying the~construction of~the~zero distribution---restricting the~survival amplitude locally to its two dominant terms---may appear overly rough.
While this approximation is indeed insufficient for even a~qualitative reproduction of~the~time dependence of~the~survival amplitude, it captures the~distribution of~zeros remarkably well.

An intuitive understanding of this fact can be gained from the~observation that the zeros of a~holomorphic function act as sources for its logarithm.
Away from zeros, $\log |\mathcal{L}(z)|$ is harmonic, while the zeros appear as singular sources in the corresponding Laplace equation.
The local variation of the survival amplitude is therefore governed primarily by the nearest zeros, which produce the strongest oscillatory features in a given region of the complex plane.
Restricting the survival amplitude locally to the two largest terms translates to picking a single dominant oscillatory frequency, generated by the corresponding chain of two-level zeros.
Additional terms then play the role of a holomorphic perturbation: they shift and deform this chain continuously, rather than eliminating it.

To estimate the~typical deformation of~a~two-level zero chain induced by additional terms, we consider a~region around $\beta_{ab}$.
The survival amplitude around $\beta_{ab}$ can be expressed as 
\begin{eqnarray}\label{eq:OverlapApproximation}
    \mathcal{L}(z) &= \ell_a(z) + \ell_b(z) + R_{ab}(\beta)e^{i\phi_{ab}(t)},
    \end{eqnarray}
    where the~third term is the~sum of~all the~remaining terms $\sum_{j \neq a,b}\ell_j(z)$.
 The~first two terms generate a~two-level zero chain while the~remaining sum acts as a~perturbation, not necessarily locally smaller than the~sum of~the~two terms.
 At longer times, when the~coherence of~the~terms is lost, we estimate the~effect of~the~additional terms by treating their sum around $z_{ab}$ as a~sum of~complex numbers with random phases.
By doing so, we neglect oscillations generated by other envelope terms, effectively treating the~associated zeros as inactive sources.
Within this approximation, the~magnitude $R_{ab}(\beta)$ follows a~Rayleigh distribution~\cite{beckmann1964rayleigh}
\begin{equation}
    p(R_{ab}) = \frac{2R_{ab}}{s_{ab}} e^{-\frac{R_{ab}^2}{s_{ab}}}, \quad
s_{ab} = \sum_{j \neq a,b} r_j^2(\beta_{ab}).
\end{equation}
The maximum of~the~distribution at $R_{ab} = \sqrt{s_{ab}}/2$ sets the~typical scale of~the~deformation, leading to the~intertwining of~neighboring zero chains.

The typical size of~the~perturbation can be expressed in~terms of~the~inverse participation ratio $\mathcal{L}_{IPR}$
of the~initial state defined in~Eq.~(\ref{eq:IPR}).
At $\beta = 0$, and for general $\beta$ after normalizing the~survival amplitude, we obtain 
\begin{equation}
    s_{ab} =\mathcal{L}_{IPR}^2 - k_a^2 - k_b^2.
\end{equation}
This reveals a~simple trend: as the~number of~populated states increases, the~typical magnitude of~the~perturbation decreases.
However, for a~fully delocalized state $k_j = 1/(d+1)$, all terms form an intersection around $\beta = 0$.
Since the~structure of~the~zero distribution is generated by variations in~the~energy distribution through the~envelope, a~flat distribution provides no organizing principle for the~zeros.
Consequently, the~distribution of~zeros cannot be decomposed into chains and instead corresponds to the~complicated zero structure of~a~canonical partition function.

\section{Energy envelope for two-band models}
\label{app:twoband}
In this appendix, we provide technical details supporting the~energy envelope interpretation of~zeros in~two-band models.
We first summarize the~Bogoliubov transformation for the~Ising model and collect explicit expressions for the~excitation amplitudes and overlaps used in~the~main text.
We then derive which eigenstates of~the~final Hamiltonian form the~envelope of~the~energy distribution for a~quenched BCS ground state.
Finally, we illustrate how multiple dominant envelope states give rise to multilevel zeros using the~XY model as an example.

\subsection{Bogoliubov transformation and overlaps of~BCS states in~the~Ising model}
The 1D transverse-field Ising model in~the~nearest-neighbor limit, $\alpha \rightarrow \infty$ in~Eq.~(\ref{LongRangeIsing}), can be mapped onto a~system of~noninteracting fermions via the~standard Jordan-Wigner and Bogoliubov transformations~\cite{10.21468/SciPostPhysLectNotes.82}. In~momentum space, the~Hamiltonian is diagonalized by the~Bogoliubov transformation
\begin{equation}
    \label{eq:Bogoliubov}
    \hat{\gamma}_q(h) = u_q(h)\hat{c}_q - v_q(h)\hat{c}^{\dagger}_{-q},
\end{equation}
where $\hat{c}_q$ are Fourier-transformed Jordan-Wigner fermions.
 
In terms of~the~new quasiparticles $\hat{\gamma}_q(h)$
, the~Hamiltonian in~the~parametrization as in~Eq.~(\ref{LongRangeIsing})
takes the~diagonal form
\begin{equation}
    \label{eq:noninteracting_ising}
    \hat{H}(h) = \frac{1}{8} \sum_{q>0} \varepsilon_q(h)\left( \gamma^{\dagger}_q\gamma_q + \gamma^{\dagger}_{-q}\gamma_{-q} - 1\right),
\end{equation}
with single-particle energies 
$$\varepsilon_q(h) = \sqrt{(4h - \cos q)^2 + \sin^2 q}.$$
The Bogoliubov coefficients are given by
    \begin{equation}
        \label{eq:coeffs}
        u_q(h) = \sqrt{\frac{\varepsilon_q(h) + s_q(h)}{2\varepsilon_q(h)}}, 
        v_q(h) = \frac{i \sin q }{2\varepsilon_q(h)(\varepsilon_q(h) + s_q(h))},
    \end{equation}
    where $s_q = 4h - \cos q$.
In this parametrization, the~model undergoes a~ground state quantum phase transition at $h_c = 1/4$.

Eigenstates of~$\hat{H}(h_i)$ and $\hat{H}(h_f)$ are related by a~Bogoliubov transformation.
For a~quench from $h_i$ to $h_f$, the~overlap between the~ground state of~the~initial Hamiltonian $|\emptyset_i\rangle$ and a~$2n$-particle eigenstate of~the~final Hamiltonian populated by pairs of~particles with allowed momenta $\{\pm q_{j_1},\dots,\pm q_{j_n}\}$ factorizes into a~product of~excitation amplitudes $Z_{-q,q}$ as given in~Eq.~(\ref{eq:GS_overlap}).
The excitation amplitudes can be expressed as
\begin{equation}
    \label{eq:zfactor}
    Z_{-q,q} = \frac{u_{-q}^{(f)} v_q^{(i)} + u_q^{(i)} v_{-q}^{(f)}}{{u_{-q}^{(i)*}} u_{-q}^{(f)} + {v_{-q}^{(i)*} } v_{-q}^{(f)}},
\end{equation}
where superscripts $(i)$ and $(f)$ denote coefficients evaluated at $h_i$ and $h_f$, respectively.

The symmetric ground state of~the~Hamiltonian in~Eq.~(\ref{eq:noninteracting_ising}) for the~Ising model belongs to the~odd sector.
Consequently, the~allowed momenta are $q \in \{\pi/N, 3\pi/N \dots (N-1)\pi/N\}$.

\subsection{Construction of~the~envelope}

In this subsection, we identify the~eigenstates of~the~final Hamiltonian that form the~envelope of~the~energy distribution for a~quenched BCS ground state, i.e., the~states satisfying the~dominance condition in~Eq.~(\ref{eq:inequality}).

We begin by ordering the~allowed momentum pairs $\pm q$ according to the~function  $W(q)$ defined in~Eq.~(\ref{eq:Wfunction}), which satisfies $W(q)=W(-q)$.
This ordering balances the~contribution of~a~quasiparticle pair to the~overlap with its associated energy cost.

Let $|\psi_a\rangle$ denote a~$2n$-particle eigenstate of~the~final Hamiltonian populated by $n$ pairs of~particles with momenta $\pm q_{j_a}$ with the~$n$ largest values of~$W(q)$.
Let $\pm q_b$ be the~pair with the~$(n+1)$-th largest value of~$W(q)$, and define the~state
 $$|\psi_b\rangle = \hat{\gamma}_{q_b}^{\dagger}\hat{\gamma}_{-q_b}^{\dagger}|\psi_a\rangle.$$

We now show that the~states $|\psi_a\rangle$ and $|\psi_b\rangle$ form consecutive elements of~the~envelope. 
The initial ground state has nonzero overlap only with eigenstates populated by quasiparticle pairs $\pm q$ from the~allowed momentum sector.
Any competing state $|\psi_c\rangle$ entering the~dominance condition in~Eq.~(\ref{eq:inequality}) can therefore be written as a~$2m$-particle state populated by a~set of~such pairs.

The set of~pairs populating $|\psi_c\rangle$ can be partitioned into two subsets: 
pairs $\pm q_{j_+}$ that also populate $|\psi_a\rangle$, and pairs $\pm q_{j_-}$ that are absent from $|\psi_a\rangle$.

By construction, the~pairs $\pm q_{j_+}$ have larger values of~$W$ than the~pair $\pm q_{b}$, satisfying the~inequality $$W(q_{j_+}) \geq W(q_{b}),$$ while for the~pairs $\pm q_{j_-}$, the~following inequality holds $$W(q_{j_-}) \leq W(q_{b}).$$ 
Consequently, the~contributions of~the~two subsets to the~overlap and to the~total energy enter the~dominance condition with opposite inequalities.
After inserting the~populations and energies of~the~states $|\psi_a\rangle$, $|\psi_b\rangle$ and $|\psi_c\rangle$, the~logarithm of~the~inequality in~Eq.~(\ref{eq:inequality}) simplifies to \begin{eqnarray*}
    -\sum_{\{q_{i+}\}} \varepsilon_{j_+} W(q_{b})
&+&\sum_{\{q_{i-}\}} \varepsilon_{j_-} W(q_{b}) 
\geq\\
-\sum_{\{q_{i+}\}} \varepsilon_{j_+} W(q_{j_+})
&+&\sum_{\{q_{i-}\}} \varepsilon_{j_-} W(q_{j_-}),
\end{eqnarray*}
which shows that the~inequality is satisfied for each momentum $q_{j_{\pm}}$ individually. 
Therefore, the~pair of~states $|\psi_a\rangle$ and $|\psi_b\rangle$ form consecutive elements of~the~envelope.

Iterating this argument demonstrates that the~envelope consists of~eigenstates obtained by successively adding pairs of~quasiparticles in~decreasing order of~$W(q)$.

\subsection{Multilevel zeros in~two-band models}
In the~example shown in~Fig.~\ref{fig:IsingMultiple} for the~nearest-neighbor Ising model, at most a~single momentum mode satisfies the~condition
$|Z_{-q,q}| = 1$.
This reflected the~monotonic behavior of~the~excitation amplitude  $|Z_{-q,q}|$ after a~quench across the~quantum critical point.
As a~result, zeros on the~real-time axis arose from isolated two-level intersections of~envelope states.

\begin{figure}
    \centering
    \includegraphics[width=1\linewidth]{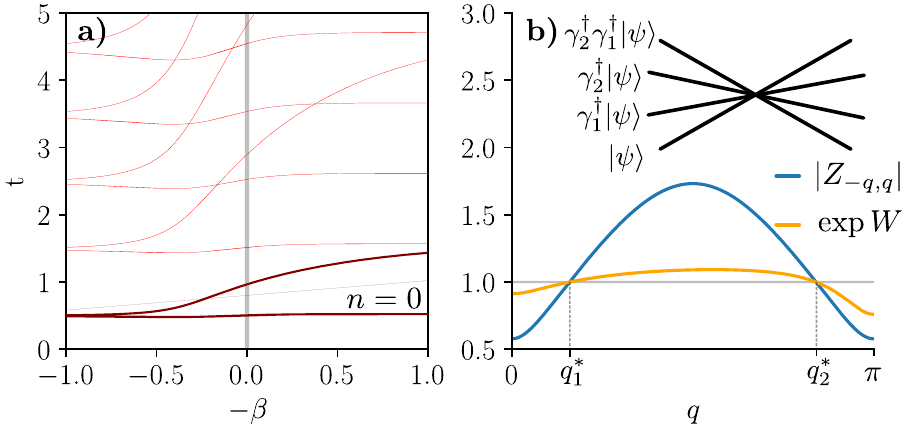}
    \caption{Lines of~zeros $z_n(q)$ and the~excitation amplitude $|Z_{-q,q}|$ for a~quench of~a~ground state in~the~XY model with parameters $\gamma_i = 1.5$, $\gamma_f = -1.5$ and $h_i = 0.5, h_f = -0.5$ as in~Ref.~\cite{PhysRevB.89.161105}.
    \textbf{(a)}  Lines of~zeros $z_n(q)$ for $n = 0,\dots 4$.
    The~line for $n = 0$ is highlighted by a~darker color to emphasize that each line crosses the~real-time axis twice.
    \textbf{(b)} the~excitation amplitude $|Z_{-q,q}|$ and the~ordering function $W(q)$ for the~quench.
    The~condition $|Z_{-q,q}| = 1$ is satisfied for two momenta $q^*_1$ and $q^*_2$.
    This distribution of~zeros along the~real-time axis corresponds to a~multilevel zero with four-level intersections of~the~radial terms corresponding to states $|\psi\rangle, \hat{\gamma}^{\dagger}_1|\psi\rangle,\hat{\gamma}^{\dagger}_2|\psi\rangle$ and $\hat{\gamma}^{\dagger}_1\hat{\gamma}^{\dagger}_2|\psi\rangle$ as shown in~the~sketch.}
    \label{fig:XYmodel}
\end{figure}

More generally, two-band models may exhibit a~more complex momentum dependence of~the~excitation amplitude, such that the~condition 
$|Z_{-q,q}| = 1$ is satisfied for multiple distinct momenta 
$q_1^{*},q_2^{*}$ \dots.
A representative example is provided by the~XY model, whose Hamiltonian with periodic boundary conditions reads
\begin{equation}
    \label{eq:HamXY}
    \hat{H}(\gamma, h) = \sum_{j=1}^{N} \left( \frac{1 + \gamma}{2} \hat{\sigma}_j^x \hat{\sigma}_{j+1}^x + \frac{1 - \gamma}{2} \hat{\sigma}_j^y \hat{\sigma}_{j+1}^y - h \hat{\sigma}_j^z \right).
\end{equation}
As shown in~Ref.~\cite{PhysRevB.89.161105}, the~XY model exhibits, in~the~thermodynamic limit, two momenta for which the~condition for the~excitation amplitude is satisfied, as illustrated in~Fig.~\ref{fig:XYmodel} (b). 
For a~finite system, an exact multilevel zero forms only when both of~these momenta coincide with available discrete momentum values.
This condition is generically fulfilled in~the~thermodynamic limit, where the~momentum spectrum becomes dense.

When this condition is met, the~distribution of~zeros of~the~survival amplitude is, for certain quenches, formed by two chains of~zeros with different periods, each corresponding to a~different momentum $q_1^*$ and $q_2^*$ as shown in~Fig.~\ref{fig:XYmodel}(a), where each line of~zeros intersects the~real-time axis twice.

If this condition were satisfied in~a~finite system, the~envelope would develop a~flat maximum formed by four equally populated states, giving rise to this special type of~multilevel distribution of~zeros.

In terms of~the~radial contributions to the~survival amplitude, all corresponding terms intersect simultaneously at a~given $\beta$ in~groups of~four, forming four-level intersections.
Each such intersection is built on an eigenstate $|\psi\rangle$ of~the~final Hamiltonian that does not contain quasiparticle pairs with momenta $\pm q_1^*$ and $\pm q_2^*$.
The four states involved in~the~intersection are
$|\psi\rangle, \hat{\gamma}^{\dagger}_1|\psi\rangle,\hat{\gamma}^{\dagger}_2|\psi\rangle$ and $\hat{\gamma}^{\dagger}_1\hat{\gamma}^{\dagger}_2|\psi\rangle$, where $\hat{\gamma}_j^{\dagger} \equiv \hat{\gamma}_{q_{j}^*}^{\dagger}\hat{\gamma}_{-q_{j}^*}^{\dagger}$, as illustrated by the~sketch in~Fig.~\ref{fig:XYmodel}(b).
The simultaneous intersections lead to exact multilevel zeros through complete cancellation in~the~survival amplitude.

\section{Analysis of~the~Gaussian model}
\label{sec:ShortTimes}
In this appendix, we present a~technical analysis of~the~zero distribution of~the~survival amplitude in~the~Gaussian model introduced in~Sec.~\ref{sec:IV}.
We first derive the~analytic form of~the~distribution of~zeros in~the~unbounded regime and then analyze the~bounded regime and derive the~trajectories of~the~zeros to first order in~the~nonuniformity parameter $\varepsilon$.

\subsection{Unbounded regime}

In the~definition of~the~Gaussian model in~Eq.~(\ref{eq:GaussianModelSurAmplitude}), the~sum runs over a~finite number of~levels such that the~level spacing remains positive.
In the~regime where the~initial state is centered deep in~the~interior of~the~spectrum, the~boundedness can be neglected and the~sum may be approximated by an infinite sum.
 
This formally yields the~elliptic theta function 
\begin{equation}
    \vartheta_3(u, \tau) \equiv \sum_{j=-\infty}^{\infty} e^{2 j i u + \pi i \tau j^2}
\end{equation}
with $u(z) = \frac{i z \Delta}{2}$ and $\tau(z) = \frac{i}{\pi}\frac{1}{2\sigma^2(z)}$.
The approximation is valid on the~half-plane $\beta \gg -\frac{1}{\varepsilon \sigma^2}$, where the~condition of~the~positive level spacing is satisfied. 

To extend the~interpretation for a~nonzero $\beta$, we normalize the~survival amplitude in~this unbounded regime
\begin{equation}
    \label{eq:NormalizedGaussianSurvivalAmplitude}
    \bar{\mathcal{L}}_U(z) = \frac{\vartheta_3(u(z),\tau(z))}{\vartheta_3(u(\beta),\tau(\beta))}, 
\end{equation}
which formally corresponds to the~survival amplitude of~an initial state with energy distribution $k_j(\beta) = k_j\exp(-E_j \beta)/{\mathcal{N}}$.

The logarithm of~the~normalized survival amplitude of~the~Gaussian model is shown in~Fig.~\ref{fig:unboundGaussian}.
The minima of~the~survival amplitude appear as peaks in~the~logarithmic plot.
Their decay is governed by dephasing, which manifests as a~shrinking of~the~effective complex-time-dependent energy width $\sigma(z)$.

This decay is accurately captured by the~function  
\begin{equation}
    \label{eq:Decay}
    \bar{\Theta}_{\text{Decay}}(z) = \frac{\vartheta_3(\frac{\pi}{2}\frac{\Delta}{\Delta_{\text{Center}}},\tau(z))}{\vartheta_3(0,\tau(\beta))} 
\end{equation}
which follows the~minima of~the~
amplitude approximately at  
$t_n = n\pi/\Delta_{\text{center}}$.
Here $2\pi/\Delta_{\text{center}}$ corresponds to the~period of~the~approximate revivals at fixed $\beta$, with
\begin{equation}
    \Delta_{\text{center}}(\beta) = \Delta + \varepsilon \frac{\beta \Delta \sigma^2}{1 + \beta \Delta \sigma^2}.
\end{equation}

For a~non-equidistant spectrum with $\varepsilon \neq 0$, the~zeros of~the~complex-time survival amplitude lie exactly at the~positions given by the~two-level formula in~Eq.~(\ref{eq:ShortTimeModel_zeros}). 
 For linearly varying level spacing $\Delta_j$, the~zeros lie on smooth curves in~the~complex plane as shown in~Fig.~\ref{fig:unboundGaussian}(b).
 These curves can be parametrized as
\begin{equation}
    \label{eq:Model_Lines}
    z_n(x) =  \frac{1}{2\sigma^2} \frac{2x -1}{\Delta + x \varepsilon} + i\pi \frac{(2n+1)}{\Delta + x \varepsilon}
\end{equation}
with the~relevant range $x> -\Delta/ \epsilon$.

The exact two-level structure of~the~zero distribution follows from the~symmetry of~the~Gaussian energy distribution, which enforces pairwise cancellations of~contributions to the~survival amplitude.
At $\beta = \beta_{j,j+1}$ not only do the~two dominant terms $\ell_{j}$ and $\ell_{j+1}$ cancel, but all remaining terms $\ell_{m}$ cancel pairwise due to the~relation  
$$r_m = r_{2j+1-m}, \quad \forall~m\in \mathbb{Z}.$$
The energy differences between paired terms are integer multiples of~the~dominant level spacing
\begin{equation}
  \frac{|\Delta_{m,2j+1-m}|}{|\Delta_{j,j+1}|}= 2j+1-2m,  
\end{equation}
which produces periodic zeros with the~period $T_{j,j+1}$ of~the~dominant pair.
Deviations from this symmetry, arising from unequal amplitudes or from shifts in~energy differences, act as perturbations that deform the~distribution of~zeros away from the~two-level form.

\subsection{Bounded regime}
Based on the~decomposition of~the~survival amplitude $\mathcal{L}_G(z)$ in~Eq.~(\ref{eq:bound_model}), we distinguish two categories of~its zeros.
The first consists of~perturbed zeros of~$\mathcal{L}_U(z)$ or $C(z)$, which occur when the~other term is smaller in~the~neighborhood of the zero.
The second consists of~zeros arising from the~condition $|\mathcal{L}_U(z)| \approx |C(z)|$, where both terms are of~comparable magnitude near the~zero.
The former are organized into nearly two-level structures, whereas the~latter are responsible for a distinct~short-time distribution that differs from the~chains in~Eq.~(\ref{eq:ShortTimeModel_zeros}).

In the~analysis of~the~zero distribution in~this regime, we first consider the~case of~constant level spacing, $\varepsilon = 0$, and treat $\varepsilon$ perturbatively afterward.
For $\varepsilon =0$, the~survival amplitude $\mathcal{L}_{G}(z)$ 
is periodic in~time with period $T = 2\pi/\Delta$.

\begin{figure}
    \centering
    \includegraphics[width=0.99\linewidth]{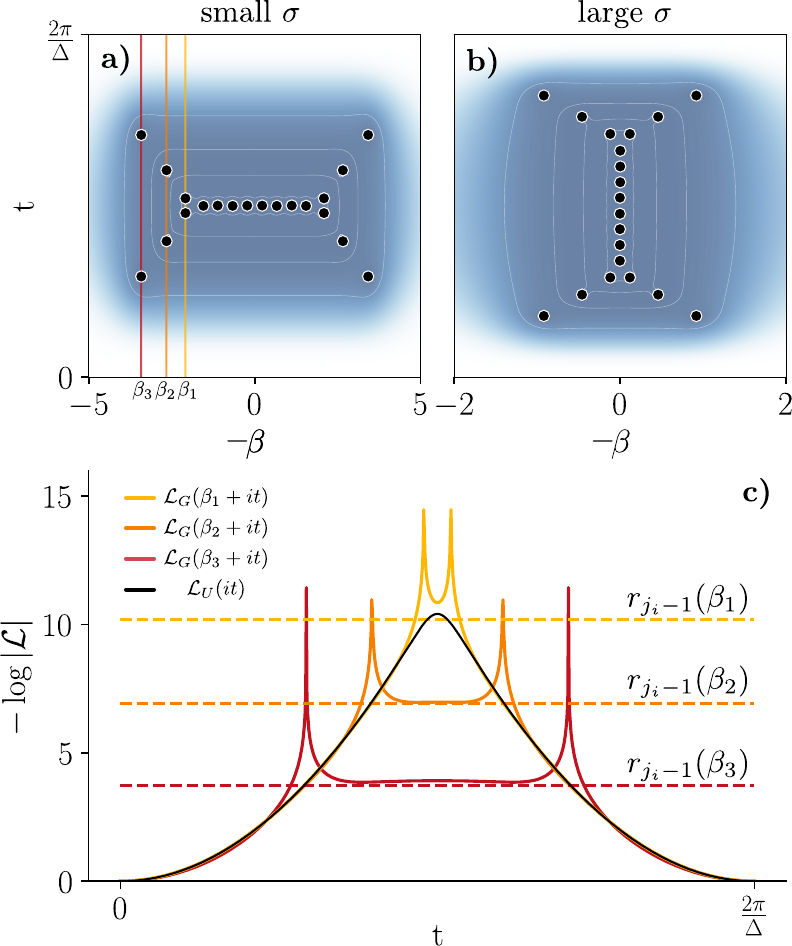}
    \caption{
    The~first row shows the~two types of~zero distribution of~the~Gaussian model $\mathcal{L}_G(z)$ for constant level spacing $\varepsilon = 0$.
    The~spectrum consists of~21 levels bounded by $j_i = -10$ and $j_f = 10$, yielding 20 zeros per period.
    Panel \textbf{(a)} corresponds to a~narrow energy distribution with $\sigma = 1.5$, while panel \textbf{(b)} shows a~broader distribution with a~width $\sigma = 2.5$.
    Panel \textbf{(c)} compares the~rate function of~the~bounded survival amplitude $\mathcal{L}_G$ at several $\beta$ values (with zeros marked by black dots) to the~unbounded term $\mathcal{L}_U$ and the~cutoff term, approximated in this region by its largest term $r_{j_i-1}$, to illustrate the~creation of~the~zeros by mutual cancellation of~these terms.
    The~highlighted pairs of~zeros at $\beta_1,\beta_2, \beta_3$ act as initial conditions for trajectories of~zeros in~Fig.~\ref{fig:runningZeros}.}    
    \label{fig:boundGaussian}
\end{figure}

Both $\mathcal{L}_U(z)$ and the~cutoff term $C(z)$ have chain-like zero structures over one period.
The unbounded survival amplitude produces horizontal chains corresponding to the~lines in~Eq.~(\ref{eq:Model_Lines}),
while the~cutoff term gives rise to vertical chains of~zeros generated by the~dominant boundary contributions $\ell_{j_i-1}$ and $\ell_{j_f+1}$.

For a~narrow width of~the~initial state, the~survival amplitude $\mathcal{L}_G(z)$ inherits the~zero distribution of~$\mathcal{L}_U(z)$.
For broader distribution extending over a~large portion of~the~spectrum, the~zero structure may instead be dominated by the~cutoff term $C(z)$.
These two regimes are illustrated in~Figs.~\ref{fig:boundGaussian}(a) and (b).

In both cases, the~wedge-shaped zeros at the~edges of~the~distribution arise from the~cancellation condition $|\mathcal{L}_U(z)| \approx |C(z)|$. 
The origin of~these zeros is illustrated in~Fig.~\ref{fig:boundGaussian}(c).
They originate from the~cancellation between the~slowly varying cutoff term, which is approximated in~the~figure by its largest term, and 
the time-dependent Gaussian peaks of~the~unbounded term $\mathcal{L}_U(z)$, located at $\pi/\Delta$ and with width proportional to $\sigma^{-1}$.
As $|\beta|$ increases, the~cutoff term grows exponentially, widening the~separation between successive zeros.
For sufficiently large $|\beta|$, the~cutoff term dominates and zeros cease to form.

For $\varepsilon \neq 0$, the~spectrum is no longer equidistant and the~distribution of~zeros is not strictly periodic.
We treat $\varepsilon$ perturbatively and determine the~shift of~the~zeros away from their positions $z_0(n)$ in~the~equidistant case.
To first order in~$\varepsilon$, the~perturbed zeros satisfy 
\begin{equation}
\label{eq:perturbationLine}
    z(\varepsilon) = z_0(n)\left( 1 - \frac{\varepsilon}{2\Delta}\mathcal{K}(z_0)\right).
\end{equation}
Here we define  
\begin{equation}
    \mathcal{K}(z) \equiv \frac{\sum_jk_jj^2\exp(-zj\Delta)}{\sum_jk_jj\exp(-zj\Delta)}.
\end{equation}
Equation~(\ref{eq:perturbationLine}) can be viewed as defining a~first-order trajectory in~the~complex plane, parametrized by $n$, with initial condition $z_0(n=0)$.

The resulting motion of~the~zeros is illustrated in~Fig.~\ref{fig:runningZeros} in~the~main text.
Starting approximately from the~zeros of~$\mathcal{L}_{G}$ with $\varepsilon = 0$ (shown in~Fig.~\ref{fig:boundGaussian}(a)), each zero follows the~trajectory given by Eq.~(\ref{eq:perturbationLine}).
In particular, zeros originating from the~wedge structure at the~same $\beta$ generally separate and move along distinct paths, reflecting the~opposite sign of~their first-order shifts.

\nocite{DQPT-I_Halimeh,marino2022dynamical}
\bibliography{bibliography}

@PREAMBLE{
 "\providecommand{\noopsort}[1]{}" 
 # "\providecommand{\singleletter}[1]{#1}%" 
}

@article{kloc2018quantum,
  title = {Quantum quench dynamics in {D}icke superradiance models},
  author = {Kloc, Michal and Str\'ansk\'y, Pavel and Cejnar, Pavel},
  journal = {Phys. Rev. A},
  volume = {98},
  issue = {1},
  pages = {013836},
  numpages = {13},
  year = {2018},
  month = {Jul},
  publisher = {American Physical Society},
  doi = {10.1103/PhysRevA.98.013836},
  url = {https://link.aps.org/doi/10.1103/PhysRevA.98.013836}
}

@article{huang2016dynamical,
  title = {Dynamical Quantum Phase Transitions: Role of Topological Nodes in Wave Function Overlaps},
  author = {Huang, Zhoushen and Balatsky, Alexander V.},
  journal = {Phys. Rev. Lett.},
  volume = {117},
  issue = {8},
  pages = {086802},
  numpages = {6},
  year = {2016},
  month = {Aug},
  publisher = {American Physical Society},
  doi = {10.1103/PhysRevLett.117.086802},
  url = {https://link.aps.org/doi/10.1103/PhysRevLett.117.086802}
}

@article{beckmann1964rayleigh,
  title={{R}ayleigh distribution and its generalizations},
  author={Beckmann, Petr},
  journal={Radio Science Journal of Research NBS/USNC-URSI},
  volume={68},
  number={9},
  pages={927--932},
  year={1964}
}

@article{heyl2013dynamical,
  title = {Dynamical Quantum Phase Transitions in the Transverse-Field {I}sing Model},
  author = {Heyl, M. and Polkovnikov, A. and Kehrein, S.},
  journal = {Phys. Rev. Lett.},
  volume = {110},
  issue = {13},
  pages = {135704},
  numpages = {5},
  year = {2013},
  month = {Mar},
  publisher = {American Physical Society},
  doi = {10.1103/PhysRevLett.110.135704},
  url = {https://link.aps.org/doi/10.1103/PhysRevLett.110.135704}
}

@article{perez2011quantum,
  title = {Quantum quench influenced by an excited-state phase transition},
  author = {P\'erez-Fern\'andez, P. and Cejnar, P. and Arias, J. M. and Dukelsky, J. and Garc\'{\i}a-Ramos, J. E. and Rela\~no, A.},
  journal = {Phys. Rev. A},
  volume = {83},
  issue = {3},
  pages = {033802},
  numpages = {14},
  year = {2011},
  month = {Mar},
  publisher = {American Physical Society},
  doi = {10.1103/PhysRevA.83.033802},
  url = {https://link.aps.org/doi/10.1103/PhysRevA.83.033802}
}

@article{cejnar2021excited,
doi = {10.1088/1751-8121/abdfe8},
url = {https://doi.org/10.1088/1751-8121/abdfe8},
year = {2021},
month = {mar},
publisher = {IOP Publishing},
volume = {54},
number = {13},
pages = {133001},
author = {Cejnar, Pavel and Stránský, Pavel and Macek, Michal and Kloc, Michal},
title = {Excited-state quantum phase transitions},
journal = {Journal of Physics A: Mathematical and Theoretical}
}

@article{botet1982size,
  title={Size scaling for infinitely coordinated systems},
  author={Botet, R and Jullien, R and Pfeuty, P},
  journal={Physical Review Letters},
  volume={49},
  number={7},
  pages={478},
  year={1982},
  publisher={APS}
}

@article{kac1963van,
  title={On the van der {W}aals Theory of the Vapor-Liquid Equilibrium. {I}. {D}iscussion of a One-Dimensional Model},
  author={Kac, M and Uhlenbeck, GE and Hemmer, PC},
  journal={Journal of Mathematical Physics},
  volume={4},
  number={2},
  pages={216--228},
  year={1963},
  publisher={American Institute of Physics}
}

@article{schmitt2015dynamical,
  title = {Dynamical quantum phase transitions in the {K}itaev honeycomb model},
  author = {Schmitt, Markus and Kehrein, Stefan},
  journal = {Phys. Rev. B},
  volume = {92},
  issue = {7},
  pages = {075114},
  numpages = {13},
  year = {2015},
  month = {Aug},
  publisher = {American Physical Society},
  doi = {10.1103/PhysRevB.92.075114},
  url = {https://link.aps.org/doi/10.1103/PhysRevB.92.075114}
}

@article{PRBKarrasch2013,
  title = {Dynamical phase transitions after quenches in nonintegrable models},
  author = {Karrasch, C. and Schuricht, D.},
  journal = {Phys. Rev. B},
  volume = {87},
  issue = {19},
  pages = {195104},
  numpages = {8},
  year = {2013},
  month = {May},
  publisher = {American Physical Society},
  doi = {10.1103/PhysRevB.87.195104},
  url = {https://link.aps.org/doi/10.1103/PhysRevB.87.195104}
}

@article{PhysRevB.89.125120,
  title = {Dynamical quantum phase transitions and the {L}oschmidt echo: A transfer matrix approach},
  author = {Andraschko, F. and Sirker, J.},
  journal = {Phys. Rev. B},
  volume = {89},
  issue = {12},
  pages = {125120},
  numpages = {13},
  year = {2014},
  month = {Mar},
  publisher = {American Physical Society},
  doi = {10.1103/PhysRevB.89.125120},
  url = {https://link.aps.org/doi/10.1103/PhysRevB.89.125120}
}

@article{PhysRevB.89.161105,
  title = {Disentangling dynamical phase transitions from equilibrium phase transitions},
  author = {Vajna, Szabolcs and D\'ora, Bal\'azs},
  journal = {Phys. Rev. B},
  volume = {89},
  issue = {16},
  pages = {161105},
  numpages = {5},
  year = {2014},
  month = {Apr},
  publisher = {American Physical Society},
  doi = {10.1103/PhysRevB.89.161105},
  url = {https://link.aps.org/doi/10.1103/PhysRevB.89.161105}
}

@article{PhysRevB.90.125106,
  title = {Dynamical quantum phase transitions in the axial next-nearest-neighbor {I}sing chain},
  author = {Kriel, J. N. and Karrasch, C. and Kehrein, S.},
  journal = {Phys. Rev. B},
  volume = {90},
  issue = {12},
  pages = {125106},
  numpages = {9},
  year = {2014},
  month = {Sep},
  publisher = {American Physical Society},
  doi = {10.1103/PhysRevB.90.125106},
  url = {https://link.aps.org/doi/10.1103/PhysRevB.90.125106}
}

@article{PhysRevB.92.104306,
  title = {Quenches and dynamical phase transitions in a nonintegrable quantum {I}sing model},
  author = {Sharma, Shraddha and Suzuki, Sei and Dutta, Amit},
  journal = {Phys. Rev. B},
  volume = {92},
  issue = {10},
  pages = {104306},
  numpages = {7},
  year = {2015},
  month = {Sep},
  publisher = {American Physical Society},
  doi = {10.1103/PhysRevB.92.104306},
  url = {https://link.aps.org/doi/10.1103/PhysRevB.92.104306}
}

@article{PhysRevB.89.054301,
  title = {Dynamical phase transitions, time-integrated observables, and geometry of states},
  author = {Hickey, James M. and Genway, Sam and Garrahan, Juan P.},
  journal = {Phys. Rev. B},
  volume = {89},
  issue = {5},
  pages = {054301},
  numpages = {9},
  year = {2014},
  month = {Feb},
  publisher = {American Physical Society},
  doi = {10.1103/PhysRevB.89.054301},
  url = {https://link.aps.org/doi/10.1103/PhysRevB.89.054301}
}

@article{PhysRevLett.119.080501,
  title = {Direct Observation of Dynamical Quantum Phase Transitions in an Interacting Many-Body System},
  author = {Jurcevic, P. and Shen, H. and Hauke, P. and Maier, C. and Brydges, T. and Hempel, C. and Lanyon, B. P. and Heyl, M. and Blatt, R. and Roos, C. F.},
  journal = {Phys. Rev. Lett.},
  volume = {119},
  issue = {8},
  pages = {080501},
  numpages = {5},
  year = {2017},
  month = {Aug},
  publisher = {American Physical Society},
  doi = {10.1103/PhysRevLett.119.080501},
  url = {https://link.aps.org/doi/10.1103/PhysRevLett.119.080501}
}

@Article{Flaschner2018,
author={Fl{\"a}schner, N.
and Vogel, D.
and Tarnowski, M.
and Rem, B. S.
and L{\"u}hmann, D.-S.
and Heyl, M.
and Budich, J. C.
and Mathey, L.
and Sengstock, K.
and Weitenberg, C.},
title={Observation of dynamical vortices after quenches in a system with topology},
journal={Nature Physics},
year={2018},
month={Mar},
day={01},
volume={14},
number={3},
pages={265-268},
issn={1745-2481},
doi={10.1038/s41567-017-0013-8},
url={https://doi.org/10.1038/s41567-017-0013-8}
}

@article{PhysRevApplied.11.044080,
  title = {Observation of a Dynamical Quantum Phase Transition by a Superconducting Qubit Simulation},
  author = {Guo, Xue-Yi and Yang, Chao and Zeng, Yu and Peng, Yi and Li, He-Kang and Deng, Hui and Jin, Yi-Rong and Chen, Shu and Zheng, Dongning and Fan, Heng},
  journal = {Phys. Rev. Appl.},
  volume = {11},
  issue = {4},
  pages = {044080},
  numpages = {12},
  year = {2019},
  month = {Apr},
  publisher = {American Physical Society},
  doi = {10.1103/PhysRevApplied.11.044080},
  url = {https://link.aps.org/doi/10.1103/PhysRevApplied.11.044080}
}

@article{PhysRevB.100.024310,
  title = {Observation of dynamical phase transitions in a topological nanomechanical system},
  author = {Tian, Tian and Ke, Yongguan and Zhang, Liang and Lin, Shaochun and Shi, Zhifu and Huang, Pu and Lee, Chaohong and Du, Jiangfeng},
  journal = {Phys. Rev. B},
  volume = {100},
  issue = {2},
  pages = {024310},
  numpages = {12},
  year = {2019},
  month = {Jul},
  publisher = {American Physical Society},
  doi = {10.1103/PhysRevB.100.024310},
  url = {https://link.aps.org/doi/10.1103/PhysRevB.100.024310}
}

@article{PhysRevLett.122.020501,
  title = {Simulating Dynamic Quantum Phase Transitions in Photonic Quantum Walks},
  author = {Wang, Kunkun and Qiu, Xingze and Xiao, Lei and Zhan, Xiang and Bian, Zhihao and Yi, Wei and Xue, Peng},
  journal = {Phys. Rev. Lett.},
  volume = {122},
  issue = {2},
  pages = {020501},
  numpages = {6},
  year = {2019},
  month = {Jan},
  publisher = {American Physical Society},
  doi = {10.1103/PhysRevLett.122.020501},
  url = {https://link.aps.org/doi/10.1103/PhysRevLett.122.020501}
}

@Article{Dborin2022,
author={Dborin, James
and Wimalaweera, Vinul
and Barratt, F.
and Ostby, Eric
and O'Brien, Thomas E.
and Green, A. G.},
title={Simulating groundstate and dynamical quantum phase transitions on a superconducting quantum computer},
journal={Nature Communications},
year={2022},
month={Oct},
day={10},
volume={13},
number={1},
pages={5977},
issn={2041-1723},
doi={10.1038/s41467-022-33737-4},
url={https://doi.org/10.1038/s41467-022-33737-4}
}

@article{PhysRevB.94.064423,
  title = {Pulse and quench induced dynamical phase transition in a chiral multiferroic spin chain},
  author = {Azimi, M. and Sekania, M. and Mishra, S. K. and Chotorlishvili, L. and Toklikishvili, Z. and Berakdar, J.},
  journal = {Phys. Rev. B},
  volume = {94},
  issue = {6},
  pages = {064423},
  numpages = {12},
  year = {2016},
  month = {Aug},
  publisher = {American Physical Society},
  doi = {10.1103/PhysRevB.94.064423},
  url = {https://link.aps.org/doi/10.1103/PhysRevB.94.064423}
}

@article{PhysRevB.99.121107,
  title = {Dynamical quantum phase transitions in collapse and revival oscillations of a quenched superfluid},
  author = {Lacki, Mateusz and Heyl, Markus},
  journal = {Phys. Rev. B},
  volume = {99},
  issue = {12},
  pages = {121107},
  numpages = {6},
  year = {2019},
  month = {Mar},
  publisher = {American Physical Society},
  doi = {10.1103/PhysRevB.99.121107},
  url = {https://link.aps.org/doi/10.1103/PhysRevB.99.121107}
}

@article{PhysRevB.99.054302,
  title = {Quench dynamics and zero-energy modes: {T}he case of the {C}reutz model},
  author = {Jafari, R. and Johannesson, Henrik and Langari, A. and Martin-Delgado, M. A.},
  journal = {Phys. Rev. B},
  volume = {99},
  issue = {5},
  pages = {054302},
  numpages = {11},
  year = {2019},
  month = {Feb},
  publisher = {American Physical Society},
  doi = {10.1103/PhysRevB.99.054302},
  url = {https://link.aps.org/doi/10.1103/PhysRevB.99.054302}
}

@article{PhysRevLett.122.250601,
  title = {Dynamical Topological Quantum Phase Transitions in Nonintegrable Models},
  author = {Hagym\'asi, I. and Hubig, C. and Legeza, \"O. and Schollw\"ock, U.},
  journal = {Phys. Rev. Lett.},
  volume = {122},
  issue = {25},
  pages = {250601},
  numpages = {6},
  year = {2019},
  month = {Jun},
  publisher = {American Physical Society},
  doi = {10.1103/PhysRevLett.122.250601},
  url = {https://link.aps.org/doi/10.1103/PhysRevLett.122.250601}
}

@article{PhysRevLett.122.050403,
  title = {Dynamical Topological Transitions in the Massive {S}chwinger Model with a $\ensuremath{\theta}$ Term},
  author = {Zache, T. V. and Mueller, N. and Schneider, J. T. and Jendrzejewski, F. and Berges, J. and Hauke, P.},
  journal = {Phys. Rev. Lett.},
  volume = {122},
  issue = {5},
  pages = {050403},
  numpages = {6},
  year = {2019},
  month = {Feb},
  publisher = {American Physical Society},
  doi = {10.1103/PhysRevLett.122.050403},
  url = {https://link.aps.org/doi/10.1103/PhysRevLett.122.050403}
}

@article{PhysRevB.101.014305,
  title = {Nonequilibrium renormalization group fixed points of the quantum clock chain and the quantum {P}otts chain},
  author = {Wu, Yantao},
  journal = {Phys. Rev. B},
  volume = {101},
  issue = {1},
  pages = {014305},
  numpages = {7},
  year = {2020},
  month = {Jan},
  publisher = {American Physical Society},
  doi = {10.1103/PhysRevB.101.014305},
  url = {https://link.aps.org/doi/10.1103/PhysRevB.101.014305}
}

@article{RYLANDS2021168554,
title = {Loschmidt echo of far-from-equilibrium fermionic superfluids},
journal = {Annals of Physics},
volume = {435},
pages = {168554},
year = {2021},
note = {Special issue on Philip W. Anderson},
issn = {0003-4916},
doi = {https://doi.org/10.1016/j.aop.2021.168554},
url = {https://www.sciencedirect.com/science/article/pii/S0003491621001603},
author = {Colin Rylands and Emil A. Yuzbashyan and Victor Gurarie and Aidan Zabalo and Victor Galitski},
}

@article{PhysRevB.103.064306,
  title = {Dynamical quantum phase transition in a bosonic system with long-range interactions},
  author = {Syed, Marvin and Enss, Tilman and Defenu, Nicol\`o},
  journal = {Phys. Rev. B},
  volume = {103},
  issue = {6},
  pages = {064306},
  numpages = {9},
  year = {2021},
  month = {Feb},
  publisher = {American Physical Society},
  doi = {10.1103/PhysRevB.103.064306},
  url = {https://link.aps.org/doi/10.1103/PhysRevB.103.064306}
}

@article{PhysRevB.104.085104,
  title = {Correlations and dynamical quantum phase transitions in an interacting topological insulator},
  author = {Yu, Wing Chi and Sacramento, P. D. and Li, Yan Chao and Lin, Hai-Qing},
  journal = {Phys. Rev. B},
  volume = {104},
  issue = {8},
  pages = {085104},
  numpages = {10},
  year = {2021},
  month = {Aug},
  publisher = {American Physical Society},
  doi = {10.1103/PhysRevB.104.085104},
  url = {https://link.aps.org/doi/10.1103/PhysRevB.104.085104}
}

@article{anomalous2017,
  title = {Dynamical phase diagram of quantum spin chains with long-range interactions},
  author = {Halimeh, Jad C. and Zauner-Stauber, Valentin},
  journal = {Phys. Rev. B},
  volume = {96},
  issue = {13},
  pages = {134427},
  numpages = {5},
  year = {2017},
  month = {Oct},
  publisher = {American Physical Society},
  doi = {10.1103/PhysRevB.96.134427},
  url = {https://link.aps.org/doi/10.1103/PhysRevB.96.134427}
}

@article{anomalousII2017,
  title = {Anomalous dynamical phase in quantum spin chains with long-range interactions},
  author = {Homrighausen, Ingo and Abeling, Nils O. and Zauner-Stauber, Valentin and Halimeh, Jad C.},
  journal = {Phys. Rev. B},
  volume = {96},
  issue = {10},
  pages = {104436},
  numpages = {7},
  year = {2017},
  month = {Sep},
  publisher = {American Physical Society},
  doi = {10.1103/PhysRevB.96.104436},
  url = {https://link.aps.org/doi/10.1103/PhysRevB.96.104436}
}

@article{PhysRevE.96.062118,
  title = {Probing the anomalous dynamical phase in long-range quantum spin chains through {F}isher-zero lines},
  author = {Zauner-Stauber, Valentin and Halimeh, Jad C.},
  journal = {Phys. Rev. E},
  volume = {96},
  issue = {6},
  pages = {062118},
  numpages = {8},
  year = {2017},
  month = {Dec},
  publisher = {American Physical Society},
  doi = {10.1103/PhysRevE.96.062118},
  url = {https://link.aps.org/doi/10.1103/PhysRevE.96.062118}
}

@article{PhysRevResearch.2.033111,
  title = {Quasiparticle origin of dynamical quantum phase transitions},
  author = {Halimeh, Jad C. and Van Damme, Maarten and Zauner-Stauber, Valentin and Vanderstraeten, Laurens},
  journal = {Phys. Rev. Res.},
  volume = {2},
  issue = {3},
  pages = {033111},
  numpages = {8},
  year = {2020},
  month = {Jul},
  publisher = {American Physical Society},
  doi = {10.1103/PhysRevResearch.2.033111},
  url = {https://link.aps.org/doi/10.1103/PhysRevResearch.2.033111}
}

@article{anomalousisnotlongrange,
  title = {Dynamical phase transitions in quantum spin models with antiferromagnetic long-range interactions},
  author = {Halimeh, Jad C. and Van Damme, Maarten and Guo, Lingzhen and Lang, Johannes and Hauke, Philipp},
  journal = {Phys. Rev. B},
  volume = {104},
  issue = {11},
  pages = {115133},
  numpages = {14},
  year = {2021},
  month = {Sep},
  publisher = {American Physical Society},
  doi = {10.1103/PhysRevB.104.115133},
  url = {https://link.aps.org/doi/10.1103/PhysRevB.104.115133}
}

@article{PhysRevB.100.014434,
  title = {Dynamical criticality and domain-wall coupling in long-range Hamiltonians},
  author = {Defenu, Nicol\`o and Enss, Tilman and Halimeh, Jad C.},
  journal = {Phys. Rev. B},
  volume = {100},
  issue = {1},
  pages = {014434},
  numpages = {10},
  year = {2019},
  month = {Jul},
  publisher = {American Physical Society},
  doi = {10.1103/PhysRevB.100.014434},
  url = {https://link.aps.org/doi/10.1103/PhysRevB.100.014434}
}

@article{PhysRevB.91.155127,
  title = {Topological classification of dynamical phase transitions},
  author = {Vajna, Szabolcs and D\'ora, Bal\'azs},
  journal = {Phys. Rev. B},
  volume = {91},
  issue = {15},
  pages = {155127},
  numpages = {5},
  year = {2015},
  month = {Apr},
  publisher = {American Physical Society},
  doi = {10.1103/PhysRevB.91.155127},
  url = {https://link.aps.org/doi/10.1103/PhysRevB.91.155127}
}

@article{PhysRevResearch.1.033039,
  title = {Experimental classification of quenched quantum walks by dynamical {C}hern number},
  author = {Xu, Xiao-Ye and Wang, Qin-Qin and Tao, Si-Jing and Pan, Wei-Wei and Chen, Zhe and Jan, Munsif and Zhan, Yong-Tao and Sun, Kai and Xu, Jin-Shi and Han, Yong-Jian and Li, Chuan-Feng and Guo, Guang-Can},
  journal = {Phys. Rev. Res.},
  volume = {1},
  issue = {3},
  pages = {033039},
  numpages = {16},
  year = {2019},
  month = {Oct},
  publisher = {American Physical Society},
  doi = {10.1103/PhysRevResearch.1.033039},
  url = {https://link.aps.org/doi/10.1103/PhysRevResearch.1.033039}
}

@article{heyl2018dynamical,
  title={Dynamical quantum phase transitions: a review},
  author={Heyl, Markus},
  journal={Reports on Progress in Physics},
  volume={81},
  number={5},
  pages={054001},
  year={2018},
  publisher={IOP Publishing}
}

@article{PhysRevB.93.085416,
  title = {Dynamical topological order parameters far from equilibrium},
  author = {Budich, Jan Carl and Heyl, Markus},
  journal = {Phys. Rev. B},
  volume = {93},
  issue = {8},
  pages = {085416},
  numpages = {7},
  year = {2016},
  month = {Feb},
  publisher = {American Physical Society},
  doi = {10.1103/PhysRevB.93.085416},
  url = {https://link.aps.org/doi/10.1103/PhysRevB.93.085416}
}

@article{PhysRevB.108.094306,
  title = {Dynamical bulk-boundary correspondence and dynamical quantum phase transitions in higher-order topological insulators},
  author = {Mas\l{}owski, T. and Sedlmayr, N.},
  journal = {Phys. Rev. B},
  volume = {108},
  issue = {9},
  pages = {094306},
  numpages = {13},
  year = {2023},
  month = {Sep},
  publisher = {American Physical Society},
  doi = {10.1103/PhysRevB.108.094306},
  url = {https://link.aps.org/doi/10.1103/PhysRevB.108.094306}
}

@article{PhysRevB.95.1843072Dtopology,
  title = {Interconnections between equilibrium topology and dynamical quantum phase transitions in a linearly ramped {H}aldane model},
  author = {Bhattacharya, Utso and Dutta, Amit},
  journal = {Phys. Rev. B},
  volume = {95},
  issue = {18},
  pages = {184307},
  numpages = {13},
  year = {2017},
  month = {May},
  publisher = {American Physical Society},
  doi = {10.1103/PhysRevB.95.184307},
  url = {https://link.aps.org/doi/10.1103/PhysRevB.95.184307}
}

@article{PhysRevB.96.014302,
  title = {Emergent topology and dynamical quantum phase transitions in two-dimensional closed quantum systems},
  author = {Bhattacharya, Utso and Dutta, Amit},
  journal = {Phys. Rev. B},
  volume = {96},
  issue = {1},
  pages = {014302},
  numpages = {6},
  year = {2017},
  month = {Jul},
  publisher = {American Physical Society},
  doi = {10.1103/PhysRevB.96.014302},
  url = {https://link.aps.org/doi/10.1103/PhysRevB.96.014302}
}

@article{DQPT-I_Halimeh,
  title = {Prethermalization and persistent order in the absence of a thermal phase transition},
  author = {Halimeh, Jad C. and Zauner-Stauber, Valentin and McCulloch, Ian P. and de Vega, In\'es and Schollw\"ock, Ulrich and Kastner, Michael},
  journal = {Phys. Rev. B},
  volume = {95},
  issue = {2},
  pages = {024302},
  numpages = {7},
  year = {2017},
  month = {Jan},
  publisher = {American Physical Society},
  doi = {10.1103/PhysRevB.95.024302},
  url = {https://link.aps.org/doi/10.1103/PhysRevB.95.024302}
}

@article{marino2022dynamical,
doi = {10.1088/1361-6633/ac906c},
url = {https://doi.org/10.1088/1361-6633/ac906c},
year = {2022},
month = {oct},
publisher = {IOP Publishing},
volume = {85},
number = {11},
pages = {116001},
author = {Marino, Jamir and Eckstein, Martin and Foster, Matthew S and Rey, Ana Maria},
title = {Dynamical phase transitions in the collisionless pre-thermal states of isolated quantum systems: theory and experiments},
journal = {Reports on Progress in Physics}
}

@article{bena2005statistical,
  title={Statistical mechanics of equilibrium and nonequilibrium phase transitions: the {Y}ang--{L}ee formalism},
  author={Bena, Ioana and Droz, Michel and Lipowski, Adam},
  journal={International Journal of Modern Physics B},
  volume={19},
  number={29},
  pages={4269--4329},
  year={2005},
  publisher={World Scientific}
}

@article{YangLee,
  title = {Statistical Theory of Equations of State and Phase Transitions. {I}. {T}heory of Condensation},
  author = {Yang, C. N. and Lee, T. D.},
  journal = {Phys. Rev.},
  volume = {87},
  issue = {3},
  pages = {404--409},
  numpages = {0},
  year = {1952},
  month = {Aug},
  publisher = {American Physical Society},
  doi = {10.1103/PhysRev.87.404},
  url = {https://link.aps.org/doi/10.1103/PhysRev.87.404}
}

@article{PhysRevLett.120.130601,
  title = {Dynamical Quantum Phase Transitions in Spin Chains with Long-Range Interactions: Merging Different Concepts of Nonequilibrium Criticality},
  author = {\ifmmode \check{Z}\else \v{Z}\fi{}unkovi\ifmmode \check{c}\else \v{c}\fi{}, Bojan and Heyl, Markus and Knap, Michael and Silva, Alessandro},
  journal = {Phys. Rev. Lett.},
  volume = {120},
  issue = {13},
  pages = {130601},
  numpages = {6},
  year = {2018},
  month = {Mar},
  publisher = {American Physical Society},
  doi = {10.1103/PhysRevLett.120.130601},
  url = {https://link.aps.org/doi/10.1103/PhysRevLett.120.130601}
}

@inproceedings{Stein2003PrincetonLI,
  title={Princeton Lectures in Analysis {II} : {C}omplex {A}nalysis},
  author={Elias M. Stein and Rami Shakarchi},
  year={2003},
  url={https://api.semanticscholar.org/CorpusID:188663800}
}

@Article{10.21468/SciPostPhysLectNotes.82,
	title={{The quantum {I}sing chain for beginners}},
	author={Glen Bigan Mbeng and Angelo Russomanno and Giuseppe E. Santoro},
	journal={SciPost Phys. Lect. Notes},
	pages={82},
	year={2024},
	publisher={SciPost},
	doi={10.21468/SciPostPhysLectNotes.82},
	url={https://scipost.org/10.21468/SciPostPhysLectNotes.82},
}

@article{PhysRevB.100.224307,
  title = {Stability of dynamical quantum phase transitions in quenched topological insulators: {F}rom multiband to disordered systems},
  author = {Mendl, Christian B. and Budich, Jan Carl},
  journal = {Phys. Rev. B},
  volume = {100},
  issue = {22},
  pages = {224307},
  numpages = {10},
  year = {2019},
  month = {Dec},
  publisher = {American Physical Society},
  doi = {10.1103/PhysRevB.100.224307},
  url = {https://link.aps.org/doi/10.1103/PhysRevB.100.224307}
}

@article{Mishra_2020,
doi = {10.1088/1751-8121/ab97de},
url = {https://doi.org/10.1088/1751-8121/ab97de},
year = {2020},
month = {aug},
publisher = {IOP Publishing},
volume = {53},
number = {37},
pages = {375301},
author = {Mishra, Utkarsh and Jafari, R and Akbari, Alireza},
title = {Disordered {K}itaev chain with long-range pairing: {L}oschmidt echo revivals and dynamical phase transitions},
journal = {Journal of Physics A: Mathematical and Theoretical},
}

@article{xu2020measuring,
  title={Measuring a dynamical topological order parameter in quantum walks},
  author={Xu, Xiao-Ye and Wang, Qin-Qin and Heyl, Markus and Budich, Jan Carl and Pan, Wei-Wei and Chen, Zhe and Jan, Munsif and Sun, Kai and Xu, Jin-Shi and Han, Yong-Jian and others},
  journal={Light: Science \& Applications},
  volume={9},
  number={1},
  pages={7},
  year={2020},
  publisher={Nature Publishing Group UK London}
}

@article{PhysRevB.101.014301,
  title = {Quasiperiodic dynamical quantum phase transitions in multiband topological insulators and connections with entanglement entropy and fidelity susceptibility},
  author = {Mas\l{}owski, T. and Sedlmayr, N.},
  journal = {Phys. Rev. B},
  volume = {101},
  issue = {1},
  pages = {014301},
  numpages = {14},
  year = {2020},
  month = {Jan},
  publisher = {American Physical Society},
  doi = {10.1103/PhysRevB.101.014301},
  url = {https://link.aps.org/doi/10.1103/PhysRevB.101.014301}
}

@article{PhysRevResearch.2.033259,
  title = {Signatures of topology in quantum quench dynamics and their interrelation},
  author = {Pastori, Lorenzo and Barbarino, Simone and Budich, Jan Carl},
  journal = {Phys. Rev. Res.},
  volume = {2},
  issue = {3},
  pages = {033259},
  numpages = {16},
  year = {2020},
  month = {Aug},
  publisher = {American Physical Society},
  doi = {10.1103/PhysRevResearch.2.033259},
  url = {https://link.aps.org/doi/10.1103/PhysRevResearch.2.033259}
}

@article{PhysRevLett.126.040602,
  title = {Entanglement View of Dynamical Quantum Phase Transitions},
  author = {De Nicola, Stefano and Michailidis, Alexios A. and Serbyn, Maksym},
  journal = {Phys. Rev. Lett.},
  volume = {126},
  issue = {4},
  pages = {040602},
  numpages = {6},
  year = {2021},
  month = {Jan},
  publisher = {American Physical Society},
  doi = {10.1103/PhysRevLett.126.040602},
  url = {https://link.aps.org/doi/10.1103/PhysRevLett.126.040602}
}

@article{PhysRevResearch.3.043064,
  title = {Mirror-symmetry-protected dynamical quantum phase transitions in topological crystalline insulators},
  author = {Okugawa, Ryo and Oshiyama, Hiroki and Ohzeki, Masayuki},
  journal = {Phys. Rev. Res.},
  volume = {3},
  issue = {4},
  pages = {043064},
  numpages = {9},
  year = {2021},
  month = {Oct},
  publisher = {American Physical Society},
  doi = {10.1103/PhysRevResearch.3.043064},
  url = {https://link.aps.org/doi/10.1103/PhysRevResearch.3.043064}
}

@article{PhysRevX.11.041018,
  title = {Determination of Dynamical Quantum Phase Transitions in Strongly Correlated Many-Body Systems Using {L}oschmidt Cumulants},
  author = {Peotta, Sebastiano and Brange, Fredrik and Deger, Aydin and Ojanen, Teemu and Flindt, Christian},
  journal = {Phys. Rev. X},
  volume = {11},
  issue = {4},
  pages = {041018},
  numpages = {15},
  year = {2021},
  month = {Oct},
  publisher = {American Physical Society},
  doi = {10.1103/PhysRevX.11.041018},
  url = {https://link.aps.org/doi/10.1103/PhysRevX.11.041018}
}

@article{PhysRevB.105.165149,
  title = {Entanglement and precession in two-dimensional dynamical quantum phase transitions},
  author = {De Nicola, Stefano and Michailidis, Alexios A. and Serbyn, Maksym},
  journal = {Phys. Rev. B},
  volume = {105},
  issue = {16},
  pages = {165149},
  numpages = {15},
  year = {2022},
  month = {Apr},
  publisher = {American Physical Society},
  doi = {10.1103/PhysRevB.105.165149},
  url = {https://link.aps.org/doi/10.1103/PhysRevB.105.165149}
}

@article{PhysRevB.100.035124,
  title = {Functional field integral approach to quantum work},
  author = {Dong, Jian-Jun and Yang, Yi-feng},
  journal = {Phys. Rev. B},
  volume = {100},
  issue = {3},
  pages = {035124},
  numpages = {11},
  year = {2019},
  month = {Jul},
  publisher = {American Physical Society},
  doi = {10.1103/PhysRevB.100.035124},
  url = {https://link.aps.org/doi/10.1103/PhysRevB.100.035124}
}

@article{PhysRevB.105.094514,
  title = {Dynamical quantum phase transitions in a mesoscopic superconducting system},
  author = {Wrze\ifmmode \acute{s}\else \'{s}\fi{}niewski, K. and Weymann, I. and Sedlmayr, N. and Doma\ifmmode \acute{n}\else \'{n}\fi{}ski, T.},
  journal = {Phys. Rev. B},
  volume = {105},
  issue = {9},
  pages = {094514},
  numpages = {6},
  year = {2022},
  month = {Mar},
  publisher = {American Physical Society},
  doi = {10.1103/PhysRevB.105.094514},
  url = {https://link.aps.org/doi/10.1103/PhysRevB.105.094514}
}

@article{PhysRevB.106.224302,
  title = {Quench dynamics and scaling laws in topological nodal loop semimetals},
  author = {Sim, Karin and Chitra, R. and Molignini, Paolo},
  journal = {Phys. Rev. B},
  volume = {106},
  issue = {22},
  pages = {224302},
  numpages = {9},
  year = {2022},
  month = {Dec},
  publisher = {American Physical Society},
  doi = {10.1103/PhysRevB.106.224302},
  url = {https://link.aps.org/doi/10.1103/PhysRevB.106.224302}
}

@article{PhysRevResearch.4.033032,
  title = {Dynamical quantum phase transitions in strongly correlated two-dimensional spin lattices following a quench},
  author = {Brange, Fredrik and Peotta, Sebastiano and Flindt, Christian and Ojanen, Teemu},
  journal = {Phys. Rev. Res.},
  volume = {4},
  issue = {3},
  pages = {033032},
  numpages = {6},
  year = {2022},
  month = {Jul},
  publisher = {American Physical Society},
  doi = {10.1103/PhysRevResearch.4.033032},
  url = {https://link.aps.org/doi/10.1103/PhysRevResearch.4.033032}
}

@article{PhysRevResearch.4.013250,
  title = {Dynamical phase transitions in the two-dimensional transverse-field {I}sing model},
  author = {Hashizume, Tomohiro and McCulloch, Ian P. and Halimeh, Jad C.},
  journal = {Phys. Rev. Res.},
  volume = {4},
  issue = {1},
  pages = {013250},
  numpages = {9},
  year = {2022},
  month = {Mar},
  publisher = {American Physical Society},
  doi = {10.1103/PhysRevResearch.4.013250},
  url = {https://link.aps.org/doi/10.1103/PhysRevResearch.4.013250}
}

@article{Maslowski_2024,
doi = {10.1088/1361-648X/ad4a16},
url = {https://doi.org/10.1088/1361-648X/ad4a16},
year = {2024},
month = {may},
publisher = {IOP Publishing},
volume = {36},
number = {33},
pages = {335401},
author = {Masłowski, Tomasz and Sedlmayr, Nicholas},
title = {The dynamical bulk boundary correspondence and dynamical quantum phase transitions in the {B}enalcazar–{B}ernevig–{H}ughes model},
journal = {Journal of Physics: Condensed Matter}
}

@article{PhysRevB.109.L140303,
  title = {Vortex loop dynamics and dynamical quantum phase transitions in three-dimensional fermion matter},
  author = {Kosior, Arkadiusz and Heyl, Markus},
  journal = {Phys. Rev. B},
  volume = {109},
  issue = {14},
  pages = {L140303},
  numpages = {6},
  year = {2024},
  month = {Apr},
  publisher = {American Physical Society},
  doi = {10.1103/PhysRevB.109.L140303},
  url = {https://link.aps.org/doi/10.1103/PhysRevB.109.L140303}
}

@article{PhysRevB.109.134301,
  title = {Distribution of {F}isher zeros in dynamical quantum phase transitions of two-dimensional topological systems},
  author = {Sacramento, P. D. and Yu, Wing Chi},
  journal = {Phys. Rev. B},
  volume = {109},
  issue = {13},
  pages = {134301},
  numpages = {18},
  year = {2024},
  month = {Apr},
  publisher = {American Physical Society},
  doi = {10.1103/PhysRevB.109.134301},
  url = {https://link.aps.org/doi/10.1103/PhysRevB.109.134301}
}

@article{Niu_2024,
doi = {10.1088/1402-4896/ad4a9d},
url = {https://doi.org/10.1088/1402-4896/ad4a9d},
year = {2024},
month = {may},
publisher = {IOP Publishing},
volume = {99},
number = {6},
pages = {065415},
author = {Niu, Zhen-Xia and Wang, Qian},
title = {Understanding dynamical phase transitions in a spinor {B}ose–{E}instein condensate via quantum and semiclassical analyses},
journal = {Physica Scripta}
}

@article{PhysRevResearch.7.023194,
  title = {Dynamical quantum phase transition and thermal equilibrium in the lattice {T}hirring model},
  author = {Ba\~nuls, Mari Carmen and Cichy, Krzysztof and Hung, Hao-Ti and Kao, Ying-Jer and Lin, C.-J. David and Singh, Amit},
  journal = {Phys. Rev. Res.},
  volume = {7},
  issue = {2},
  pages = {023194},
  numpages = {13},
  year = {2025},
  month = {May},
  publisher = {American Physical Society},
  doi = {10.1103/PhysRevResearch.7.023194},
  url = {https://link.aps.org/doi/10.1103/PhysRevResearch.7.023194}
}

@article{PhysRevB.106.045410,
  title = {Nonlinear current and dynamical quantum phase transitions in the flux-quenched {S}u-{S}chrieffer-{H}eeger model},
  author = {Rossi, Lorenzo and Dolcini, Fabrizio},
  journal = {Phys. Rev. B},
  volume = {106},
  issue = {4},
  pages = {045410},
  numpages = {8},
  year = {2022},
  month = {Jul},
  publisher = {American Physical Society},
  doi = {10.1103/PhysRevB.106.045410},
  url = {https://link.aps.org/doi/10.1103/PhysRevB.106.045410}
}

@article{Cao_2024,
doi = {10.1088/1361-648X/ad1a5a},
url = {https://doi.org/10.1088/1361-648X/ad1a5a},
year = {2024},
month = {jan},
publisher = {IOP Publishing},
volume = {36},
number = {15},
pages = {155401},
author = {Cao, Kaiyuan and Guo, Hao and Yang, Guangwen},
title = {Aperiodic dynamical quantum phase transition in multi-band {B}loch Hamiltonian and its origin},
journal = {Journal of Physics: Condensed Matter}
}

@article{PhysRevA.97.053621,
  title = {Dynamical quantum phase transitions in discrete time crystals},
  author = {Kosior, Arkadiusz and Sacha, Krzysztof},
  journal = {Phys. Rev. A},
  volume = {97},
  issue = {5},
  pages = {053621},
  numpages = {8},
  year = {2018},
  month = {May},
  publisher = {American Physical Society},
  doi = {10.1103/PhysRevA.97.053621},
  url = {https://link.aps.org/doi/10.1103/PhysRevA.97.053621}
}

@article{PhysRevB.100.085308,
  title = {Floquet dynamical quantum phase transitions},
  author = {Yang, Kai and Zhou, Longwen and Ma, Wenchao and Kong, Xi and Wang, Pengfei and Qin, Xi and Rong, Xing and Wang, Ya and Shi, Fazhan and Gong, Jiangbin and Du, Jiangfeng},
  journal = {Phys. Rev. B},
  volume = {100},
  issue = {8},
  pages = {085308},
  numpages = {11},
  year = {2019},
  month = {Aug},
  publisher = {American Physical Society},
  doi = {10.1103/PhysRevB.100.085308},
  url = {https://link.aps.org/doi/10.1103/PhysRevB.100.085308}
}

@article{PhysRevA.98.022129,
  title = {Dynamical quantum phase transitions in non-{H}ermitian lattices},
  author = {Zhou, Longwen and Wang, Qing-hai and Wang, Hailong and Gong, Jiangbin},
  journal = {Phys. Rev. A},
  volume = {98},
  issue = {2},
  pages = {022129},
  numpages = {15},
  year = {2018},
  month = {Aug},
  publisher = {American Physical Society},
  doi = {10.1103/PhysRevA.98.022129},
  url = {https://link.aps.org/doi/10.1103/PhysRevA.98.022129}
}

@article{PhysRevB.93.104302,
  title = {Quantum quench dynamics in the transverse field {I}sing model at nonzero temperatures},
  author = {Abeling, Nils O. and Kehrein, Stefan},
  journal = {Phys. Rev. B},
  volume = {93},
  issue = {10},
  pages = {104302},
  numpages = {10},
  year = {2016},
  month = {Mar},
  publisher = {American Physical Society},
  doi = {10.1103/PhysRevB.93.104302},
  url = {https://link.aps.org/doi/10.1103/PhysRevB.93.104302}
}

@article{PhysRevB.97.045147,
  title = {Fate of dynamical phase transitions at finite temperatures and in open systems},
  author = {Sedlmayr, N. and Fleischhauer, M. and Sirker, J.},
  journal = {Phys. Rev. B},
  volume = {97},
  issue = {4},
  pages = {045147},
  numpages = {8},
  year = {2018},
  month = {Jan},
  publisher = {American Physical Society},
  doi = {10.1103/PhysRevB.97.045147},
  url = {https://link.aps.org/doi/10.1103/PhysRevB.97.045147}
}

@article{Lerma-Hernandez_2018,
doi = {10.1088/1751-8121/aae2c3},
url = {https://doi.org/10.1088/1751-8121/aae2c3},
year = {2018},
month = {oct},
publisher = {IOP Publishing},
volume = {51},
number = {47},
pages = {475302},
author = {Lerma-Hernández, Sergio and Chávez-Carlos, Jorge and Bastarrachea-Magnani, Miguel A and Santos, Lea F and Hirsch, Jorge G},
title = {Analytical description of the survival probability of coherent states in regular regimes},
journal = {Journal of Physics A: Mathematical and Theoretical}
}

@article{PhysRevResearch.5.033090,
  title = {Anatomy of dynamical quantum phase transitions},
  author = {Van Damme, Maarten and Desaules, Jean-Yves and Papi\ifmmode \acute{c}\else \'{c}\fi{}, Zlatko and Halimeh, Jad C.},
  journal = {Phys. Rev. Res.},
  volume = {5},
  issue = {3},
  pages = {033090},
  numpages = {7},
  year = {2023},
  month = {Aug},
  publisher = {American Physical Society},
  doi = {10.1103/PhysRevResearch.5.033090},
  url = {https://link.aps.org/doi/10.1103/PhysRevResearch.5.033090}
}

@article{PhysRevA.103.032213,
  title = {Quasiclassical approach to quantum quench dynamics in the presence of an excited-state quantum phase transition},
  author = {Kloc, Michal and \ifmmode \check{S}\else \v{S}\fi{}imsa, Daniel and Han\'ak, Filip and Kapr\'alov\'a-\ifmmode \check{Z}\else \v{Z}\fi{}\ifmmode \check{d}\else \v{d}\fi{}\'ansk\'a, Petra Ruth and Str\'ansk\'y, Pavel and Cejnar, Pavel},
  journal = {Phys. Rev. A},
  volume = {103},
  issue = {3},
  pages = {032213},
  numpages = {14},
  year = {2021},
  month = {Mar},
  publisher = {American Physical Society},
  doi = {10.1103/PhysRevA.103.032213},
  url = {https://link.aps.org/doi/10.1103/PhysRevA.103.032213}
}

@article{PhysRevA.104.053722,
  title = {Stabilization of product states and excited-state quantum phase transitions in a coupled qubit-field system},
  author = {Str\'ansk\'y, Pavel and Cejnar, Pavel and Filip, Radim},
  journal = {Phys. Rev. A},
  volume = {104},
  issue = {5},
  pages = {053722},
  numpages = {14},
  year = {2021},
  month = {Nov},
  publisher = {American Physical Society},
  doi = {10.1103/PhysRevA.104.053722},
  url = {https://link.aps.org/doi/10.1103/PhysRevA.104.053722}
}

@article{PhysRevB.107.094307,
  title = {Mechanism of dynamical phase transitions: The complex-time survival amplitude},
  author = {Corps, \'Angel L. and Str\'ansk\'y, Pavel and Cejnar, Pavel},
  journal = {Phys. Rev. B},
  volume = {107},
  issue = {9},
  pages = {094307},
  numpages = {15},
  year = {2023},
  month = {Mar},
  publisher = {American Physical Society},
  doi = {10.1103/PhysRevB.107.094307},
  url = {https://link.aps.org/doi/10.1103/PhysRevB.107.094307}
}

@article{PhysRevB.106.024311,
  title = {Dynamical and excited-state quantum phase transitions in collective systems},
  author = {Corps, \'Angel L. and Rela\~no, Armando},
  journal = {Phys. Rev. B},
  volume = {106},
  issue = {2},
  pages = {024311},
  numpages = {21},
  year = {2022},
  month = {Jul},
  publisher = {American Physical Society},
  doi = {10.1103/PhysRevB.106.024311},
  url = {https://link.aps.org/doi/10.1103/PhysRevB.106.024311}
}

@software{jan_strelecek_2026_19734887,
  author       = {Jan Střeleček},
  title        = {strelda/ComplexifiedFidelityAnalysis: Code for
                   finding zeros of the complex-time survival
                   amplitudes
                  },
  month        = apr,
  year         = 2026,
  publisher    = {Zenodo},
  version      = {v1},
  doi          = {10.5281/zenodo.19734887},
  url          = {https://doi.org/10.5281/zenodo.19734887},
  swhid        = {swh:1:dir:ef20cc2daf5cfad7c9fb98dd54da3a5c32f60836
                   ;origin=https://doi.org/10.5281/zenodo.19734886;vi
                   sit=swh:1:snp:5a9da35cad5b6b881bfc76b34774c182bbda
                   9a61;anchor=swh:1:rel:8cd0f7c861ac03efcbb1203a85af
                   6b5b55123702;path=strelda-
                   ComplexifiedFidelityAnalysis-995c2e5
                  },
}

@incollection{Fisher1965,
  author    = {Fisher, M. E.},
  booktitle = {Boulder Lectures in Theoretical Physics},
  publisher = {University of Colorado},
  address   = {Boulder},
  year      = {1965},
  volume    = {7}
}

\end{document}